\documentclass[fleqn,usenatbib]{mnras}

\usepackage{mathptmx}
\usepackage[T1]{fontenc}
\usepackage{ae,aecompl}

\usepackage{graphicx}	% Including figure files
\usepackage{amsmath}	% Advanced maths commands
\usepackage{amssymb}	% Extra maths symbols

\usepackage{xspace}
\usepackage{bm}
\usepackage{subcaption}
\usepackage[dvipsnames]{xcolor}
\usepackage[titletoc]{appendix}

\newcommand{\ha}{H{\sc\,i}\xspace}
\newcommand{\htwo}{H{\sc\,ii}\xspace}
\newcommand{\ie}{i.e.\xspace}

\definecolor{lblue}{rgb}{0.1,0.7,1.}
\definecolor{grey}{rgb}{0.75,0.75,0.75}
\definecolor{Orange}{rgb}{1.0,0.5,0.15}

\title[Multi-wavelength consensus of large-scale linear bias]{Multi-wavelength consensus of large-scale linear bias}

\author[H. Pan et al.]{Hengxing Pan$^{1,2,3,4}$\thanks{E-mail: hengxing.pan@physics.ox.ac.uk},
Danail Obreschkow$^{1,5}$,
Cullan Howlett$^{1}$,
Claudia del P. Lagos$^{1,5,8}$,
\newauthor
Pascal J. Elahi$^{1,5}$,
Carlton Baugh$^{6}$,
Violeta Gonzalez-Perez$^{7}$\\
$^{1}$International Centre for Radio Astronomy Research, University of Western Australia, 35 Stirling Hwy, Crawley, WA 6009, Australia\\
$^{2}$Purple Mountain Observatory, 2 West beijing Road, Nanjing 210008, China\\
$^{3}$University of the Chinese Academy of Sciences, 19A, Yuquan Road, Beijing 100049, China\\
$^{4}$Astrophysics, University of Oxford, Denys wilkinson Buiding, Keble Road, Oxford OX1 3RH, UK\\
$^{5}$ARC Centre of Excellence for All Sky Astrophysics in 3 Dimensions (ASTRO 3D)\\
$^{6}$Institute for Computational Cosmology, Department of Physics, Durham University, Durham DH1 3LE, UK\\
$^{7}$Astrophysics Research Institute, Liverpool John Moores University, 146 Brownlow Hill, Liverpool L3 5RF, UK\\
$^{8}$Cosmic Dawn Center (DAWN), Niels Bohr Institute, University of Copenhagen, Copenhagen, Denmark
}

\date{Accepted XXX. Received YYY; in original form ZZZ}

\pubyear{2018}

\begin{document}
\label{firstpage}
\pagerange{\pageref{firstpage}--\pageref{lastpage}}
\maketitle

\begin{abstract}
We model the large-scale linear galaxy bias $b_g(x,z)$ as a function of redshift $z$ and observed absolute magnitude threshold $x$ for broadband continuum emission from the far infrared to ultra-violet, as well as for prominent emission lines, such as the H$\alpha$, H$\beta$, Lya and [OII] lines. 
The modelling relies on the semi-analytic galaxy formation model \textsc{GALFORM}, run on the state-of-the-art $N$-body simulation \textsc{SURFS} with the Planck 2015 cosmology. 
We find that both the differential bias at observed absolute magnitude $x$ and the cumulative bias for magnitudes brighter than $x$ can be fitted with a five-parameter model: $b_g(x,z)=a + b(1+z)^e(1 + \exp{[(x-c)d]})$.
We also find that the bias for the continuum bands follows a very similar form regardless of wavelength due to the mixing of star-forming and quiescent galaxies in a magnitude limited survey. Differences in bias only become apparent when an additional colour separation is included, which suggest extensions to this work could look at different colours at fixed magnitude limits. We test our fitting formula against observations, finding reasonable agreement with some measurements within $1\sigma$ statistical uncertainties, and highlighting areas of improvement.
We provide the fitting parameters for various continuum bands, emission lines and intrinsic galaxy properties, enabling a quick estimation of the linear bias in any typical survey of large-scale structure.

\end{abstract}

\begin{keywords}
large-scale bias, galaxy surveys, galaxy formation
\end{keywords}

%%%%%%%%%%%%%%%%%%%%%%%%%%%%%%%%%%%%%%%%%%%%%%%%%%

%%%%%%%%%%%%%%%%% BODY OF PAPER %%%%%%%%%%%%%%%%%%
      %---------------------------------------------------------------%
      %                                                               %
      %                        INTRODUCTION                           %
      %                                                               %
      %---------------------------------------------------------------%

\section{INTRODUCTION}
\label{sec:introduction}
% Paragraph
Most surveys of the cosmic large-scale structure (LSS) rely on galaxies as tracers of the dark matter distribution. 
In order to extract the cosmological information from such surveys, it is crucial to understand the difference between the spatial statistics of the detectable galaxies and the underlying density field. At the largest scales, this difference takes the form of a scaling factor, known as linear galaxy bias $b_g$ \citep{1984ApJ...284L...9K}, between the matter and galaxy power spectrum. Prior knowledge of this bias is essential in designing cosmological surveys. 

Moreover, the galaxy bias has an important role in the redshift space distortions (RSD) and multi-tracer analyses as a nuisance variable in the former case \citep[e.g.][]{2015clerkin, 2018li} and an integral part of target selection in the latter \citep[e.g.][]{2009mcdonald, 2016raul}. Predicting the galaxy bias from simulations can be used to place priors on galaxy bias in RSD measurements, and compared to lensing and LSS measurements to improve our understanding of galaxy physics. Knowing the galaxy bias can also help to break the classic degeneracy between the linear bias and cosmological parameters such as $\sigma_8$ (the linear rms of the dark matter density perturbations on scales of 8 $h^{-1}$Mpc) and the linear
growth rate $f$ present in 2-point statistics (e.g. \citealt{Ali2018} and references therein); hence priors on the bias might help better determine these parameters.   

% Paragraph
It is possible to measure the galaxy bias by computing the ratio of the two point correlation function and three point correlation function \citep[see][and references therein]{2002MNRAS.335..432V,2005MNRAS.364..620G} or by exploiting phase-space correlations \citep{Ali2018}. The bias can further be constrained by combining galaxy redshift surveys with gravitational lensing data \citep[e.g.][]{simon2007gabods,jullo2012cosmos}. Existing bias measurements correspond to specific survey selection criteria, such as selecting the galaxies seen in the near infrared (IR) band \citep{2010MNRAS.405.1006O}, the H$\alpha$ emission line \citep{amendola2017constraints}, or 21 cm emission from neutral hydrogen (\ha) \citep{2017MNRAS.471.1788C}.  However, these bias measurements are often uncertain and their extrapolation to different surveys, \ie different wavelengths and sensitivities, is challenged by the complex radiative physics of galaxies. For instance, ongoing and upcoming LSS surveys (BOSS\footnote{\url{https://www.sdss3.org/surveys/boss.php}}, eBOSS\footnote{\url{https://www.sdss.org/surveys/eboss}}, DESI\footnote{\url{https://www.desi.lbl.gov}}, LSST\footnote{\url{https://www.lsst.org}}, 4MOST\footnote{\url{https://www.4most.eu/cms}}, EUCLID\footnote{\url{https://www.euclid-ec.org}} for which the references are shown in Section~\ref{sec:results}) are covering a wide range of wavelengths and many of them also use specific emission lines to probe the cosmic LSS. Predictive bias models would help calibrating the analysis of the expected data from these surveys. In addition, it might also be interesting to explore the galaxy bias dependence on some physical properties, such as star formation rate or stellar age, in order to better understand the physical origin of specific bias values. Providing a consensus of the linear bias for various broadband wavelengths, emission lines and physical properties, is therefore a pressing goal.

% Paragraph
On the theoretical side, there are two classes of approaches to model the clustering and bias of galaxies \citep{2013PASA...30...30B}. The first class consists of populating simulated or analytically evolved dark matter halos with galaxies drawn from observed luminosity functions, for instance using the so-called halo occupation distribution (HOD) \citep[e.g.][]{berlind2002halo,zheng2005theoretical,wechsler2018connection} method or subhalo abundance matching (SHAM) \citep[e.g.][]{vale2004linking,shankar2006new}. The second class uses hydrodynamics simulations \citep[e.g.][]{schaye2014eagle,vogelsberger2014properties} and semi-analytic models \citep[SAMs, e.g.][]{2016MNRAS.462.3854L, henriques2015galaxy, lagos2018shark, croton2016semi, somerville2015star, xie2017h2} to directly model the bayonic physics involved in galaxy formation. A limitation of the first approach compared to the second is that it is descriptive rather than predictive. Gas dynamics simulations are currently limited to relatively small volumes with large shot noise on linear bias \citep{2007MNRAS.374.1479G}. To address clustering on scales of hundreds of Mpc, semi-analytics models are therefore a sensible choice. The results can in return serve as useful reference points for HOD/SHAM analyses \citep[e.g.][]{2016guo, 2016MNRAS.460.3100C}, as well as for comparisons with hydro simulations such as EAGLE \citep{2017crain} and IllustrisTNG \citep{2018springel}, which would aid in building a better understanding of the broad features of galaxy evolution. 

% Paragraph
In this paper, we use the semi-analytic model \textsc{GALFORM} \citep{cole2000hierarchical}, specifically an updated variant of \cite{2018MNRAS.474.4024G}  to model the linear galaxy bias in the standard $\mathrm{\Lambda CDM}$ cosmology with Planck (2015) parameters \citep{ade2016planck}, for various IR/optical/UV bands, emission lines and intrinsic galaxy properties. 

% Paragraph
This paper is organised as follows. In Section~\ref{sec:Methods}, we describe our dark matter $N$-body simulation with a semi-analytic galaxy formation model, the techniques used for measuring the large-scale biases and a heuristic 5-parameter model for fitting. In Section~\ref{sec:results}, we show the bias dependence on galaxy properties, emission lines and continuum bands, following up with the large-scale galaxy bias as a function of wavelength. In Section~\ref{sec:discussions}, we discuss the comparison of our model with existing surveys and forecasts, as well as the limitations of this work. In Section~\ref{sec:conclusions}, we conclude with a short synopsis of the paper.

%---------------------------------------------------------------%
%                                                               %
%                     SIMULATIONS AND METHODS                   %
%                                                               %
%---------------------------------------------------------------%
\section{SIMULATIONS AND METHODS}
\label{sec:Methods}
\subsection{Simulations and galaxy formation model}
\label{sec:Simulations}
% Paragraph
The \textsc{SURFS} suite consists of N-body simulations of a periodic volume of side lengths from 40 $h^{-1}$Mpc to 900 $h^{-1}$Mpc assuming a Planck (2015) cosmology, where $h$ is the dimensionless Hubble parameter. The simulation used in this work, L210N1536, has a side length of 210 $h^{-1}$Mpc and number of dark matter particles of $1536^3$. This choice of parameters allows us to resolve the host halos of galaxies with stellar masses of $10^8 h^{-1}\rm M_\odot$ at z = 0, with the nominal requirement that the host dark matter halos of such galaxies are resolved with 100 particles, necessary to get accountable halo masses, positions and velocities. This simulation has been run with a memory lean version of the \textsc{GADGET2} code on the Magnus supercomputer at the Pawsey supercomputing centre. For a detailed description of the simulations refer to \cite{2018MNRAS.475.5338E}, and to \cite{poulton2018observing} for a description and performance demonstration of the merger trees used here. 

\begin{figure}
	\includegraphics[width=\columnwidth]{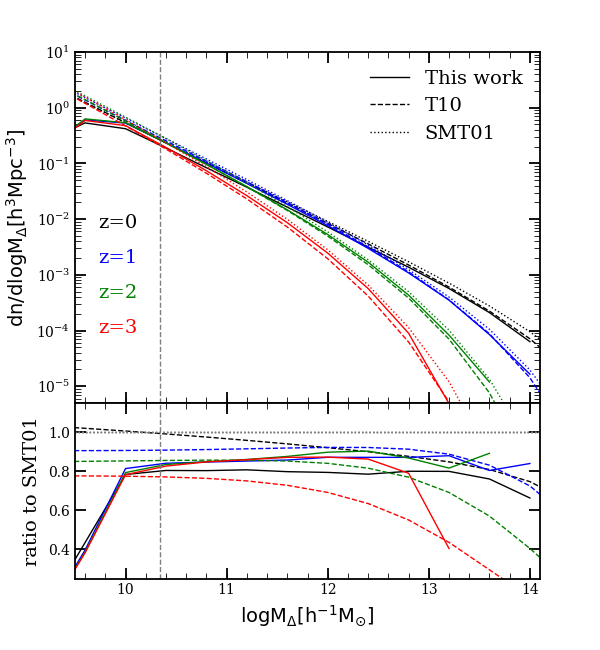}
    \caption{Halo mass function at four redshifts. 
            The solid lines are the measurements from the L210N1536 SURFS simulation. 
            The dotted and dashed lines show the SMT10 and T10 prediction respectively, calculated using \textsc{HMFcalc} \citep{2013A&C.....3...23M}.
            The gray dashed vertical line shows the halo mass limit $2.2\times 10^{10} h^{-1}\rm M_\odot$.
            Note that we use the same color caption for redshifts 
            in all bias plots as this one unless captioned differently.
            The lower panel shows the residuals relative to SMT01 model.
            }
    \label{fig:hmf}
\end{figure}

% Paragraph
The halo and subhalo catalogs are constructed by \textsc{Velociraptor} \citep{elahi2011peaks}. This code first identifies halos using a 3D Friends-of-Friends(FOF) algorithm in configuration space and then identifies subhalos using a 6D phase-space FOF algorithm on particles that differ dynamically from the dark matter background. We run \textsc{Dhalos} in the SURFS halo catalogs to adapt \textsc{VELOCIraptor} outputs to \textsc{GALFORM} inputs. \textsc{Dhalo} is a tool developed to produce and clean merger trees, which form the basis of \textsc{GALFORM} \citep[see][]{2014MNRAS.440.2115J}. We compare our halo mass function with previous works by using the virial mass defined as $M_\Delta = 4\pi R_\Delta^3 \Delta \rho_{\rm crit}/3$, $\Delta = 200$, where $\rho_{\rm crit}$ is the critical density of the universe and $R_\Delta$ is the radius that encloses this mass. In Fig.~\ref{fig:hmf}, we present the evolving halo mass function compared with \cite{sheth2001ellipsoidal} (hereafter SMT01) and \cite{2010ApJ...724..878T} (hereafter T10) models in the upper panel and the residuals relative to SMT01 in the lower panel. As shown, we find good agreement with T10 model, with an average value of uncertainties less than $10\%$ above the halo mass limit (2.2$\times 10^{10} h^{-1} \rm M_\odot$). The deviation from the SMT01 model is due to the difference of definition of halo mass as explained in \cite{tinker2008toward} and T10.

% Paragraph
Galaxy formation is approximated as a two-stage process: structure forms by hierarchical clustering in the dark matter and baryons then fall into the gravitational potential wells to form galaxies by gas cooling, star formation, feedback and stellar evolution \citep{cole2000hierarchical} %
The \textsc{GALFORM} model explicitly accounts for: 1) the shock-heating and radiative cooling of gas inside dark matter halos that drive the formation of gaseous galactic disks; 2) star formation in galaxy disks and in bulges (i.e. starbursts); 3) the growth of super massive black holes and feedback from supernovae, active galactic nucleus (AGN) as well as photo-ionization of the intergalactic medium (IGM); 4) galaxy mergers and bar instabilities which can drive bursts of star formation and lead to the formation of spheroids; 5) calculation of the sizes of disks and spheroids; 6) chemical enrichment of stars and gas; 7) calculation of galaxy stellar luminosities from the predicted star formation and chemical enrichment histories of a stellar population synthesis model; 8) nebular emission line luminosities and equivalent widths; 9) dust attenuation. We use the version of \textsc{GALFORM} based on \citet{2018MNRAS.474.4024G}, hereafter GP18, to investigate the large-scale galaxy biases. This model assumes a single initial mass function (IMF) building upon the previous versions \citep{gonzalez2014sensitive}, which is the major difference from another widely used \textsc{GALFORM} version \citep[][hereafter L16]{2016MNRAS.462.3854L}. The GP18 semi-analytical model has incorporated the merger scheme used in \cite{simha2017modelling} and the gradual stripping of hot gas when satellite galaxies are merging into central galaxies \citep{lagos2014galaxies}.

% Paragraph
We specified this model with a set of free parameters which were chosen to provide a reasonable match to the K-band luminosity function from z = 2 to z = 0 and the $b_j$ luminosity function at z = 0. We adopt the photoionisation model (Eq. 5) of \cite{kim2015h} with reionisation parameters of a circular velocity cut-off $v_{\rm cut} = 50$ km/s at a redshift of reionization $z_{\rm cut} = 10$ and the fitting parameter $\alpha_v =-0.82$ in the notation of \citet{2014MNRAS.440.1662S}. We also take the star formation efficiency parameter of the molecular gas $\nu_{\rm sf} = 0.8 Gyr^{-1}$ (Eq. 7 in L16) and the ratio of cooling/free-fall time $\alpha_{\rm cool} = 0.7$ (Eq. 12 in L16) which is an adjustable AGN feedback parameter (more galaxies to be affected by AGN feedback with larger values). The relevant figures used for the adjustment of parameters are shown in Appendix~\ref{appendixobs} and the undiscussed parameters are the same as used in \citet{2018MNRAS.474.4024G}.

% Paragraph
The model divides the baryons into five different components: hot gas for cooling in halos, a reservoir of gas ejected by feedback processes, cold gas, stars and central black holes in galaxies. 
We assumed that the galaxies have separate disk and spheroid components, which can both contain stars and cold gas. We split the cold gas into atomic hydrogen and molecular hydrogen \citep{lagos2011cosmic}; this distinction is explicitly made throughout the model at every time step. The accreted gas from the halo is added to the disk. The subsequent galaxy mergers and disk instabilities can transfer the gas into a starburst component in the spheroid. Thus we assume two separate modes of star formation: the quiescent mode (in the disk) and the starburst mode (in the spheroid). We calculate the star formation rate in the disk from the molecular gas using the empirical relation in \cite{blitz2006role} , which is based on observations of nearby star-forming disk galaxies as described in \cite{lagos2011impact}. For star formation in bursts, we assume all of the cold gas is molecular. Here we compute the SFR bias including both the quiescent and starburst modes. We also calculate the stellar mass bias by including all the stars in the galaxy.

% Paragraph
Broad-band luminosities and absolute magnitudes are calculated from the stellar SEDs of galaxies using a stellar population synthesis model (L16) based on stellar evolution models. This model also includes a simple model for emission lines in star-forming galaxies that uses the number of ionizing photons and the metallicity of the cold, star forming gas to predict emission line luminosities based on the properties of a typical \htwo region \citep{stasinska1990grid} (see GP18 for an expanded discussion of the modelling of emission lines). 

To simulate the effects of dust extinction, \textsc{GALFORM} applies a physical model of absorption and emission of radiation by dust. The dust is assumed to be present in two components: diffuse dust (75\%) and molecular clouds (25\%)  based on observations of nearby galaxies \citep{granato2000infrared}. This model includes a self-consistent model for the reprocessing of starlight by dust, in which the UV, optical and near-IR light are absorbed by dusts and reradiated at IR and sub-mm wavelengths. The dust absorption is based on radiative transfer and the temperature of the dust emission can be solved for by energy balance (see L16 for details). 

Redshifted magnitudes are needed for predicting the linear bias of future surveys. With the intrinsic properties of galaxies and their cosmological redshifts, we are able to evaluate observer-frame absolute magnitudes by
\begin{equation}
    M = -2.5\log\left[\frac{\int L_v(v_e)R(v_o)dv_e}{L_{v_o}\int R(v_o)dv_e}\right],
	\label{eq:obsmag}
\end{equation}
where $L_v(v_e)$ is the emitted luminosity per unit frequency and $L_{v_o}$ is the reference luminosity. The emitted (rest-frame) frequency $v_e$ is related to the observed (observer-frame) frequency $v_o$ by $v_e = v_o(1+z)$ and $R(v_o)$ is the filter response of a specified photometric band on the observer frame \citep[Eq. 13 in][]{merson2012lightcone}. Note that we use the symbol $M$ without subscripts as the absolute magnitude to differentiate it from the mass symbol with subscripts. The effect of dust attenuation is included throughout this paper. 

% Paragraph
For the galaxy samples chosen by continuum filters, we use the observer frame quantities to investigate the bias dependence. The UV/optical/IR filters corresponds to those used in the GALEX/SDSS/UKIRT surveys, except that the Y-band filter is from the UKIDSS survey.
% Figure

% Paragraph
%---------------------------------------------------------------%
%                                                               %
%                        Large scale bias                       %
%                                                               %
%---------------------------------------------------------------%
\subsection{Large-scale bias}
\label{sec:Large scale bias}
% Paragraph
Based on the current galaxy formation models, galaxies are formed in dark halos, therefore the understanding of halo bias is an essential component of any theory of galaxy bias \citep{smith2007scale}.

%                                                               %
%                         	halo bias  	                        %
%                                                               %
\subsubsection{Halo bias}
% Paragraph
The halo bias is determined by the relative abundance of
halos in different large-scale density environments. There are two well-known models to describe the halo bias, the peak bias \citep{bardeen1986statistics} and peak-background split model \citep{1984ApJ...284L...9K,cole1989biased,1996MNRAS.282..347M}. The peak bias model is built upon the simplified assumption that collapsed structures form from peaks in the Gaussian initial density field. Given a uniform density background at the early age of the universe, the peak bias model works well till the large-scale structures emerge and act like a local modification of the background density. The more general peak-background split model decomposes the density field into a long-wavelength and short-wavelength part. These models and the numerical calibration in T10 have given us valuable insights into the physics of the halo bias. The most obvious result is the strong dependence of large-scale halo bias on halo mass and redshift. Of course, there are other halo variables that can also produce strong trends even at fixed halo mass, such as the local tidal environment \citep[e.g.][]{2018MNRAS.476.3631P}.

% Paragraph
We review some basic theories for the halo bias. The two central quantities are the dark matter overdensity, $\delta$, and halo overdensity, $\delta_h$, 
\begin{equation}
    \delta=\frac{\rho-\bar{\rho}}{\bar{\rho}},\delta_h=\frac{n_h-\bar{n}_h}{\bar{n}_h},
	\label{eq:delta}
\end{equation}
where $\rho$ and $\bar{\rho}$ are the dark matter density and the mean density respectively, and $n_h$ and $\bar{n}_h$ are the halo number density and its mean number density respectively. 

% Paragraph
The halo bias is essentially a relation between $\delta$ and $\delta_h$, which can be Taylor expanded to \citep{fry1993biasing,mo1997high,pollack2012modelling}
\begin{equation}
    \delta_h=b_0+b_1\delta+\frac{b_2}{2}\delta^2+...,
	\label{eq:bias}
\end{equation}
where the halo bias is assumed to be local and deterministic. On large scales, it is commonly truncated to first order and the relation simply becomes linear, where $b_0 = 0$ owing to the fact that $\langle\delta_h\rangle \equiv \langle\delta\rangle \equiv 0$ by definition \citep{fry1993biasing}. These assumptions leave $b_1$ as the only relevant parameter and it is commonly regarded as the large-scale halo bias. We note that the halo bias should in principle also contain terms describing non-locality and stochasticity which become important at quasi-linear scales \citep[see Eq. 2.135 in][]{2016arXiv161109787D}, but we are focusing only on measuring the linear bias here.
% and we rewrite it as $b_h$ for denoting the large-scale halo bias . 

% Paragraph
By looking at equation \eqref{eq:bias} in Fourier space, we can also define a practical measurement of halo bias via the  relation between the halo and matter power spectrum
\begin{equation}
    b_h(k)=\sqrt{\frac{P_{h}(k)}{P_{m}(k)}},
	\label{eq:bh}
\end{equation}
where $P_{m}(k)=\langle|\delta(\bm{k})|^2\rangle$ is the dark matter power spectrum, $\delta(\bm{k})$ is the Fourier representation of dark matter overdensities, and $P_{h}(k)$ accordingly is the halo power spectrum. On large scales, it is easy to see that equation \eqref{eq:bh} can be derived from equation \eqref{eq:bias}, which means $b_h(k) = b_1$ when the wavenumber $k$ (i.e. the magnitude of the wave vector $\bm{k}$) is small in Fourier space. Hereafter we denote the large-scale halo bias as $b_h$ and specify the measurements in Section~\ref{sec:measurement}. 

%                                                               %
%                         galaxy bias                           %
%                                                               %
\subsubsection{Galaxy bias}
% Paragraph
Galaxies do not trace the mass in the same way as halos do. The difference is that galaxy formation proceeds with an efficiency which depends on halo properties in a non-linear fashion. In the lowest mass halos, feedback from supernovae and UV background ionization prevent efficient star formation  while in the high mass halos gas is unable to cool efficiently \citep{2000MNRAS.311..793B} as is heated by AGN \citep{bower2006breaking}. A comprehensive overview of galaxy bias based on perturbation theory can be seen in \cite{2016arXiv161109787D}. Following  the  definition  of halo  bias,  the galaxy bias can be written as
\begin{equation}
    b_g(k)=\sqrt{\frac{P_{g}(k)}{P_{m}(k)}},
	\label{eq:bg}
\end{equation}
where $P_{g}(k)$ is the galaxy power spectrum. 

% Paragraph
Statistically speaking, the galaxy bias can also be expressed via the halo model \citep[see][for a review]{cooray2002halo}. According to this model, the distribution of galaxies depends on how they populated halos. Hence, we can split the galaxy power spectrum into two components, the 1- and 2-halo terms: $P_{g}(k) = P_{g}^{1h}(k) + P_{g}^{2h}(k)$. The 1-halo term $P_{g}^{1h}(k)$ is determined by the density profiles of galaxy-pairs in shared halos and the halo mass function. The 2-halo term $P_{g}^{2h}(k)$ contains the contribution of the galaxies in different halos.
% Figure
\begin{figure*}
    \includegraphics[width=\textwidth]{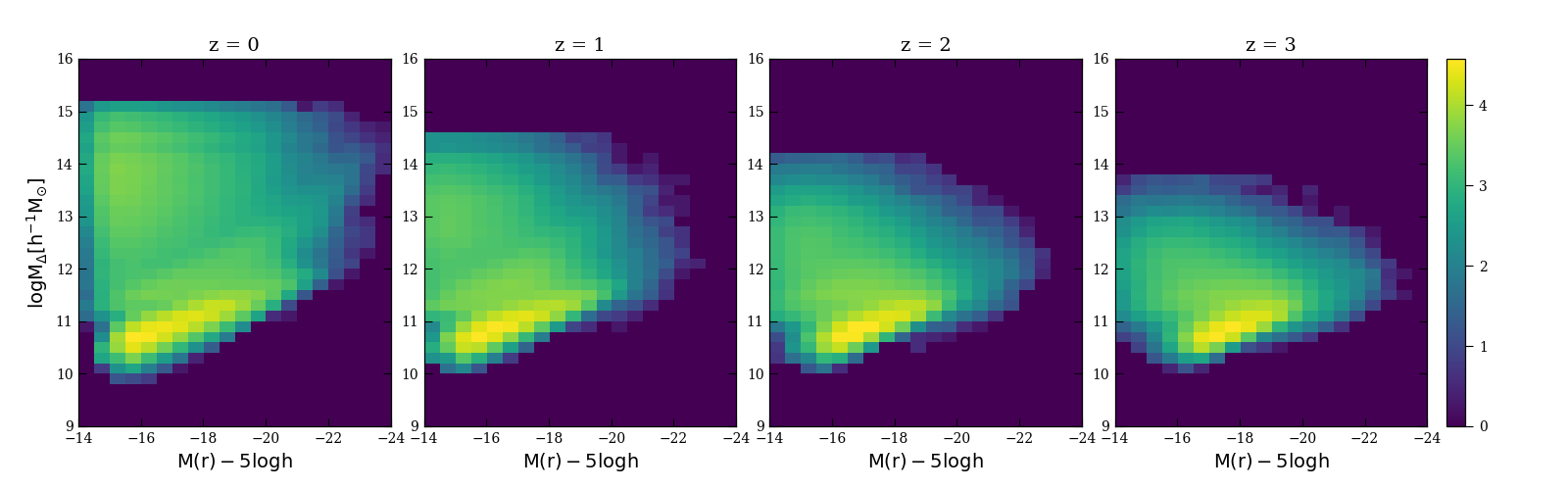}
    \caption{Two-dimensional distribution function of galaxies as a function of r-band luminosity and halo mass.
    		Results are shown at $z =  0, 1, 2, 3$ from left panel to right panel. The color map gives the log of the galaxy number.
            }
    \label{fig:hvm}
\end{figure*}

% Paragraph
On large scales, only the 2-halo term is important and it can be linked with the halo power spectrum if one knows the galaxy distribution at given mass halos. In this case, we can also rewrite the equation~\eqref{eq:bg} by weighting the large-scale halo bias with the number of galaxies contained in the given mass halos \citep[see][for the detailed derivation]{mo2010galaxy}
\begin{equation}
    b_g(x)=\frac{\int b_h(M_h)\phi(x|M_h)n(M_h)dM_h}{\int \phi(x|M_h)n(M_h)dM_h},
	\label{eq:dif}
\end{equation}
where $n(M_h)$ is the halo mass function and $\phi(x|M_h)$ denotes the conditional galaxy distribution function, such as conditional galaxy mass, star formation rate and luminosity function. The $\phi(x|M_h)$ describes the distribution of galaxies in halos of a given halo of mass $M_h$, which can be computed by the conditional PDF of host halo mass against galaxy property x from Fig.~\ref{fig:hvm}. The looming bimodal distribution from z = 3 to z = 0 is contributed by the central galaxies and satellites due to increased merger events. Here $b_g(x)$ is the large-scale differential galaxy bias, which satisfies the observational selection criteria of galaxy property $x$. Different properties $x$ will be discussed in Section \ref{sec:model}. We note that this equation emphasizes that the galaxy bias only depends on halo mass (i.e. zero galaxy assembly bias), which in practice may not be the case as indicated by many recent studies \citep[e.g.][]{2014MNRAS.443.3044Z, 2019MNRAS.484.1133C}.

% Paragraph
Following the same spirit, we note $b_g(\geq x)$ to be the large-scale cumulative galaxy bias above a limited value of $x$, which can be expressed as follows
\begin{equation}
    b_g(\geq x)=\frac{\int_{x'=x}^{+\infty} b_g(x')\phi(x')dx'}{\int_{x'=x}^{+\infty}\phi(x')dx'},
	\label{eq:cum}
\end{equation}
where $\phi(x')$ is the distribution function for a galaxy property, such as a luminosity.

%                                                               %
%    			Measurements of bias and uncertainty            %
%                                                               %
\subsection{Measurements of bias and uncertainty}
\label{sec:measurement}
% Paragraph
To measure the bias from equation \eqref{eq:bh} and \eqref{eq:bg}, one needs to calculate the power spectrum from the halo and galaxy catalogs. We first binned the halos and galaxies into a $256^3$ density mesh by using a top-hat smoothing and computed the Fourier transform to evaluate the power spectrum. All the power spectra are shot-noise subtracted and we divide each Fourier mode by a sinc-function to correct for the top-hat gridding effect. Finally we average the orientation-dependent power spectrum over a spherical shell in k-space in order to take out the orientation dependence and get $P_h(k)$ and $P_g(k)$.

% Paragraph
Since we have only one simulation to produce the galaxy samples, we rely on the theory of Gaussian covariance matrices of the power spectrum to measure the uncertainty on the bias. The commonly used expression for the power spectrum variance was derived by \cite{1994ApJ...426...23F,1997PhRvL..79.3806T}:
\begin{equation}
    \sigma_{P_g(k)}=\sqrt{\frac{2}{n_{\rm modes}}}\left[P_g(k)+\frac{1}{\bar{n}_g}\right],
	\label{eq:error_ps}
\end{equation}
where $\bar{n}_g$ refers to the number density of galaxies which satisfy the given selection criteria and $n_{\rm modes}=V4\pi{k^2}\delta{k}/(2\pi)^3$ is the number of Fourier modes in the spherical shell of width $\delta{k}$ when the volume $V \gg (2\pi/k)^3$. Finally, we provide the uncertainties of large-scale galaxy bias through error propagation as follows
\begin{equation}
	\sigma_{b_g(k)} \approx \sqrt[]{\frac{1}{n_{\rm modes}\bar{n}_g P_{m}(k)}}.
	\label{eq:berror}
\end{equation}
A derivation and verification with simulations can be found in Appendix~\ref{appendixunc}. We also employ the commonly-used "jackknife" method to estimate the error on the bias in comparison with this derived form as discussed later.

\begin{figure*}
\centering
  \begin{subfigure}{\columnwidth}
    \centering
    \includegraphics[width=\columnwidth]{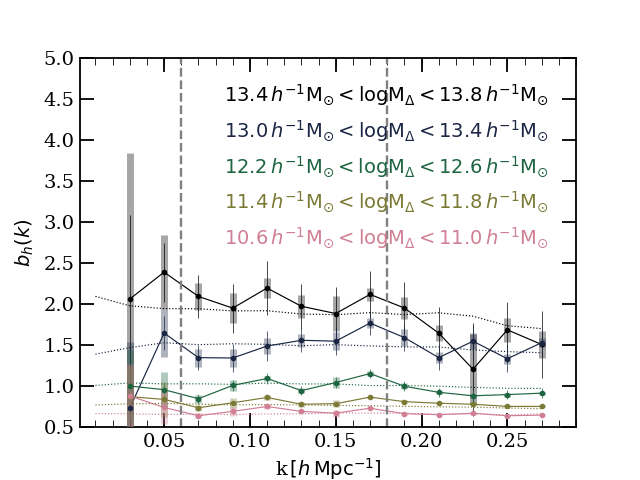}
    \label{fig:sub1}
  \end{subfigure}%
  \begin{subfigure}{\columnwidth}
    \centering
    \includegraphics[width=\columnwidth]{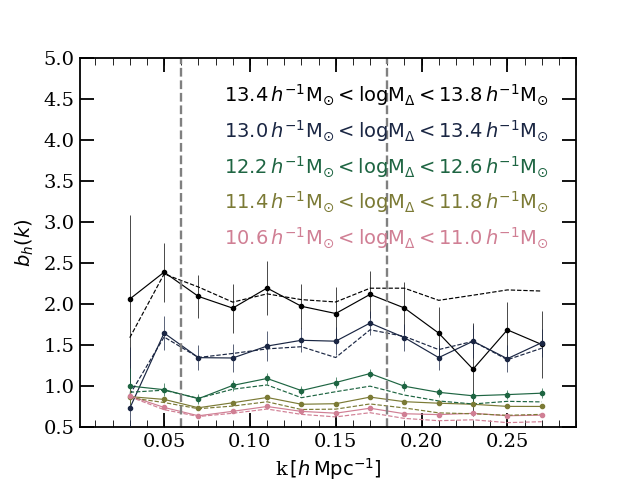}
    \label{fig:sub2}
  \end{subfigure}
    \caption{Large-scale halo bias as a function of wavenumber $k$ at z = 0.33. 
            Left: the solid and dotted lines are the measurements from the L210N1536 and L900N2048 SURFS simulations, respectively, using equation \eqref{eq:bh}. The narrower solid errors are estimated by equation \eqref{eq:berror} while the wider shaded errors are calculated by the "jackknife" method. 
            Right: the dashed lines are measured by the cross-power spectrum between halo and dark matter over the dark matter power spectrum from the L210N1536 simulation while the solid lines are the same as on the left panel. 
            The gray dashed vertical lines indicate the lower and upper limits of $2\times2\pi/L_{\rm box}$ and $0.18 h^{-1}$Mpc in $k$ space, where the $L_{\rm box} = 210 h^{-1}$Mpc.
            Only five mass bins are shown for clarity. 
            The mass bin width is 0.4 in the logarithmic space.}
    \label{fig:halobias}
  \label{fig:gr}
\end{figure*}

% Paragraph
Fig. \ref{fig:halobias} shows the halo bias as a function of wavenumber $k$ for halos of varying mass at $z = 0.33$. The solid and dotted lines are measured from the L210N1536 and L900N2048 SURFS simulations respectively using the equation \eqref{eq:bh} on the left panel. As the linear bias is independent of scale, one can see that the halo bias generally stays constant for $2\times2\pi/L_{\rm box} < k < 0.18 h^{-1}$Mpc and even applicable to larger $k$ in the low mass bins for the L210 simulation, thus we can directly measure the large-scale halo bias by averaging $b_h(k)$ over those modes. 
At higher redshifts, these limits on $k$ are more conservative since the nonlinear scale keeps extending towards the smaller $k$ as the universe evolves. Comparing the halo bias of L210 simulation with that of L900 simulation, we are confident that the L210 simulation can be used to study the large-scale bias for a variety of halo mass ranges, except for the most massive samples due to the low number statistics. The shaded error bars for the L210 simulation are estimated by the "jackknife" method. To do so, we calculate the power spectrum for eight subsamples with removing one octant of the box in each subsample, then estimate the errors on the $b_h(k)$ as indicated by the shaded area. The equation~\eqref{eq:berror} are generally consistent with this jackknife method for the less massive halos ($<10^{13} h^{-1} \rm M_\odot$), but tends to overestimate the uncertainties for the massive samples by a factor of two in the considered $k$ scales, which makes the equation~\eqref{eq:berror} more conservative. Therefore we will carry on with the derived form and exclude the points with relative errors larger than 20\% in all the bias plots. 
% Figure

We also check the halo bias as defined by the cross-power spectrum between halo and dark matter over the dark matter power spectrum indicated by the dashed lines on the right panel of Fig. \ref{fig:halobias}. As seen, this measure, whilst immune to shot-noise effects, could be contaminated with higher order bias on larger $k$ scales than the definition of the equation \eqref{eq:bh} for the low mass halos and this is in line with the findings in \citet{smith2007} and \citet{vlah2013}. Considering that the effects of shot-noise are outside the considered $k$ scales (i.e. between the dashed vertical lines) , we therefore chose to use equations \eqref{eq:bh} and \eqref{eq:bg} for the following analyses on the large-scale biases.

%Paragraph
We then average the $b_g(k)$ over small $k$ scales weighted by number of modes, which gives the large-scale galaxy bias as
\begin{equation} 
	b_g = \frac{\sum_k w_k b_g(k)}{\sum_k w_k}.
    \label{eq:bga}
\end{equation}
The variance of this weighted average is given by
\begin{equation} 
	\sigma_{b_g}^2 = \frac{\sum_k w_k (b_g(k) - b_g)^2}{\sum_k w_k},
    \label{eq:sbga}
\end{equation}
where $w_k = 1/\sigma_{b_g(k)}^2 $ are the weights given to each measurement.

\begin{figure}
    \includegraphics[width=\columnwidth]{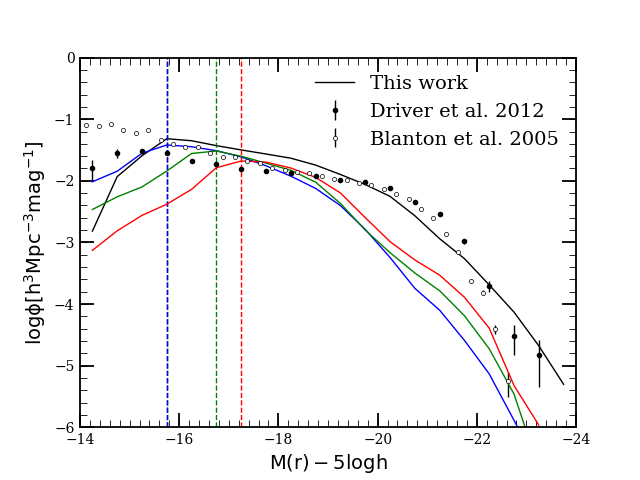}
    \caption{The r-band galaxy luminosity function at redshifts z = 0 (black), 1 (blue), 2 (green), 3 (red). 
             The dashed vertical lines show the peaks of these functions, which were used to define the resolution limit of the model.
             The black dots and open circles indicate the r-band luminosity function measured by \citet{driver2012galaxy} and \citet{blanton2005properties} at z = 0.
             }
    \label{fig:glf}
\end{figure}

% Paragraph
The galaxy bias depends on the selection of galaxies. Using $x$ as a general placeholder for a scalar galaxy property (e.g. a luminosity, an emission line strength or the stellar mass), the bias $b_g(x)$ and the cumulative bias $b_g(\geq x)$ can only be determined down to a certain value $x_{\rm min}$, below which the mass-resolution of the simulation is insufficient. To determine $x_{\rm min}$ for each property, we first apply a stellar mass cut of $10^8 h^{-1}\rm M_\odot$ to all galaxies. Above this limit, the the model is roughly complete in stellar mass. We then plot the space density function of $x$ (e.g. the luminosity function if $x$ is a luminosity) and determine the point where this function peaks. Finally, we compare the peak positions at the four redshifts (z = 0, 1, 2, 3) and set $x_{\rm min}$ equal to the largest peak value. For instance, Fig.~\ref{fig:glf} shows the galaxy r-band luminosity function compared with observations. The vertical dashed lines indicate the resolution limit for each redshift. The red dashed line represents the magnitude of -17.2, which is the brightest value compared with that of other redshifts, therefore we take it as the lower limit of r-band bias plot. We apply this same principle to all other galaxy bias plots as shown in Section \ref{sec:results}. Note that we use the stellar mass cut only for determining the lower limits of biases, not for excluding the samples at given selection criteria.
% Figure

% Paragraph
We have thus introduced two methods to compute the large-scale galaxy bias, either through averaging equation \eqref{eq:bg}, i.e. equation \eqref{eq:bga} or through the halo model, i.e. equation \eqref{eq:dif} and \eqref{eq:cum}. For clarification, we refer these two methods as "Measurement" and "Halo model" in the upcoming bias plots. 

\subsection{Heuristic 5-parameter model}
\label{sec:model}
% Figure
\begin{figure}
	\includegraphics[width=\columnwidth]{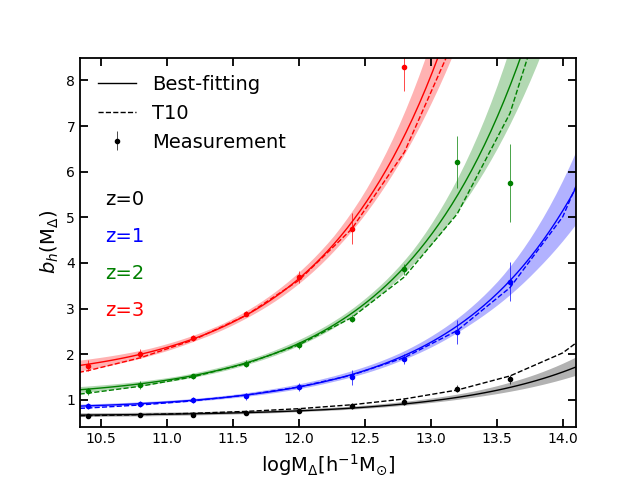}
	\includegraphics[width=\columnwidth]{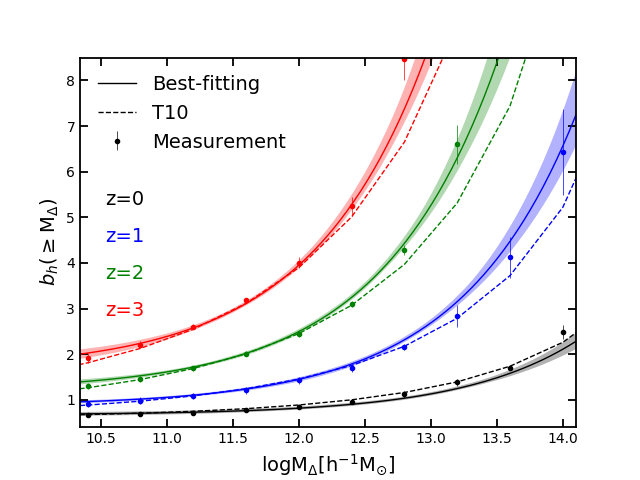}
    \caption{Large-scale halo bias as a function of halo mass at four redshifts. 
            The upper and lower panels show the differential and 
            cumulative biases respectively. 
            The filled symbols are directly measured by averaging equation~\eqref{eq:bg} on large scales. Only measurements with  a relative uncertainty below than 20\% are shown.
            The error bars are estimated by linearly propagating the shot noise uncertainties of the power spectrum.
            Solid lines show the best fits of equation~\eqref{eq:fit} and 
            dashed lines show the T10 results where we utilise the equation~\eqref{eq:cum} to compute the T10 prediction for the cumulative halo bias in the lower panel.
            The color-coded regions are the 68\% credible intervals in the halo bias estimated from the posterior samples.
            Note that we use the 3D defined halo mass here to compare with the T10.
            }
    \label{fig:bh}
\end{figure}

\begin{figure}
	\includegraphics[width=\columnwidth]{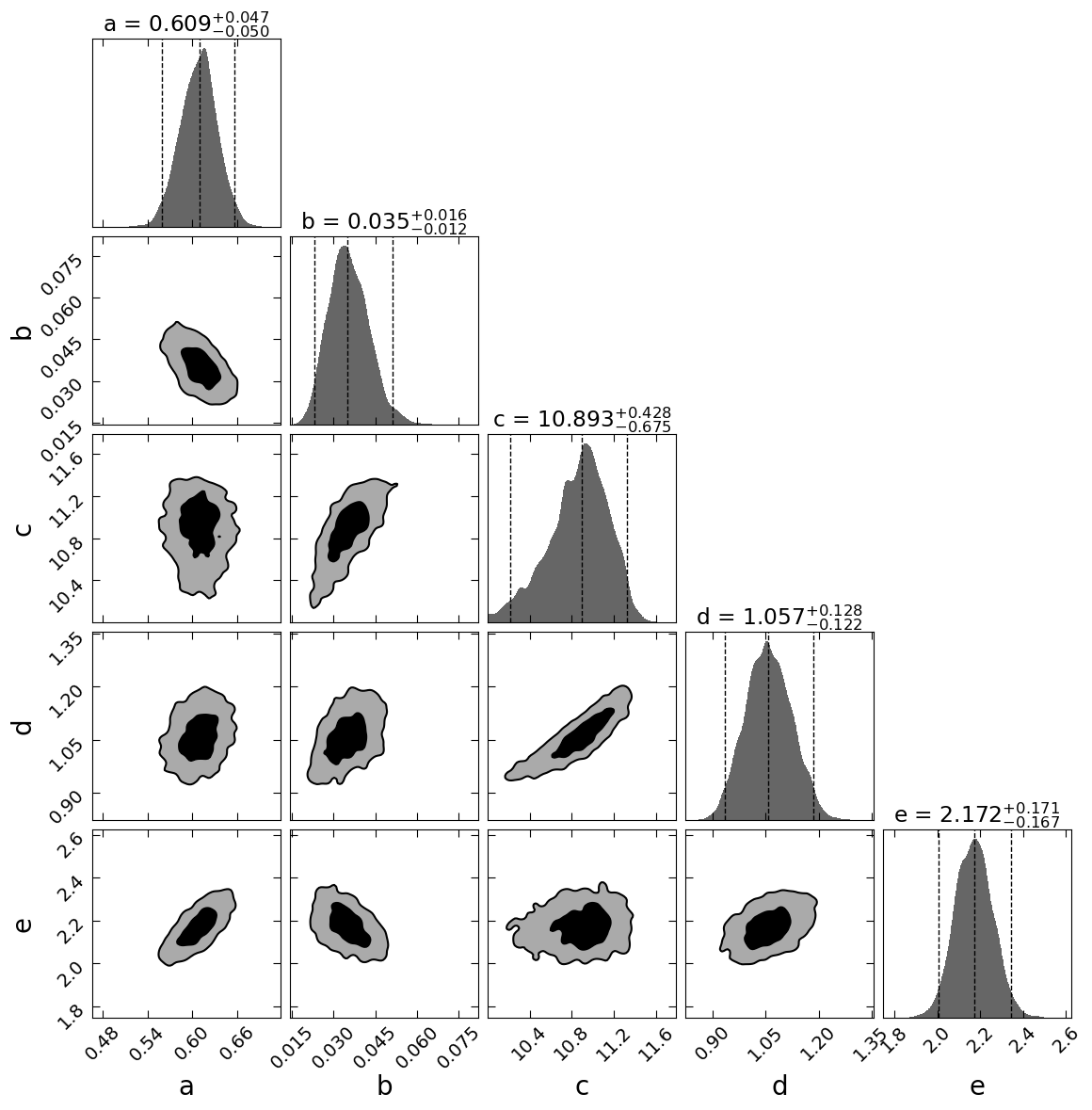}
    \caption{The posterior distributions for the 5 parameters of equation~\eqref{eq:fit} for the large-scale differential halo bias. The contour levels in the 2-D marginalized posteriors are 1 and 2$\sigma$ while the dashed lines in the 1-D marginalized posteriors span the 2$\sigma$(95\%) credible interval. }
    \label{fig:bh_post}
\end{figure}

% Paragraph
Previous bias models in the literature usually concentrate on a specific galaxy property, such as the galaxy clustering dependence on galaxy color in a single-band filter \citep{zehavi2011galaxy}, or the redshift dependence \citep{2015MNRAS.448.1389C}. The T10 model uses a universal fitting function which accurately accounts for the mass, redshift and cosmology dependence of halo bias, but cannot easily be linked to observable galaxy properties. 

% Paragraph
To improve on these models, we now fit different biases using the 5-parameter model. 
\begin{equation}
    b_g(x,z)=a + b(1+z)^e\left(1 + \exp{[(x-c)d]}\right),
	\label{eq:fit}
\end{equation}
where x is a galaxy (or halo) property, z is the redshift and $a$, $b$, $c$, $d$, and $e$ are the five parameters to be fitted. The combination of $a$, $b$, and $e$ acts as a normalization, whereas $c$ and $d$ represent the upturn point and slope on the high ends capturing the major differences of galaxy biases. We use this same formula for fitting the differential bias and the cumulative bias. If $x$ denote a magnitude (defined as negative log of the luminosity), we replace $x-c$ in equation~\eqref{eq:fit} by $c-x$ to cope with the fact the smaller magnitudes correspond to brighter objects. 

% Paragraph
Assuming that the distribution of errors follows a Gaussian, we can write the likelihood function as:
\begin{equation}
    \ln P(b_g(x,z)| a, b, c, d, e) = -\frac{1}{2}\sum \bigg[ \frac{(b_g(x,z)-b_g)^2}{\sigma_{b_g}^2} +\ln(2\pi\sigma_{b_g}^2)\bigg],
    \label{eq:maxl}
\end{equation}
where $b_g$ is the large-scale bias measured from the equation~\eqref{eq:bga} and $\sigma_{b_g}$ is the corresponding error measured from equation~\eqref{eq:sbga}. We use {\sc Multinest}\footnote{\url{https://github.com/JohannesBuchner/PyMultiNest}} to fit our model to the "Measurement" and accept the median of posterior samples as the best estimate of each parameter. Multinest is an efficient and robust Bayesian inference tool based on a nested sampling technique \citep{skilling2004nested}, which allows model fitting and produces the posterior samples. 

% Paragraph
We test our measurements and fitting formula against the well-calibrated large-scale halo bias model by T10. Fig.~\ref{fig:bh} shows the large-scale halo bias as a function of halo mass at four redshifts. Overall, our measurements for the differential halo bias show great agreement with the T10 model. The deviation on the high mass end at high redshifts is due to the low number statistics in the simulation box with a side length of 210 $h^{-1}$Mpc. It is clear that our five-parameter fitting formula can do as fine a job as the six-parameter model does in T10, except for the high mass end at redshift 0. The notable difference between our measurement and T10 prediction on the high mass ends in the lower panel indicates their differences on the halo mass function. The posterior distributions of the five fitted parameters for the large-scale differential halo bias are shown in Fig.~\ref{fig:bh_post}, demonstrating that these parameters are unimodal and well converged.

% Paragraph
The best-fitting parameters for all the selected samples can be seen from Table \ref{params_db} and \ref{params_cb} for the differential and cumulative biases respectively, which show the differential and cumulative biases are almost identical with the maximum percentage of difference less than $20\%$ . The similarity between the differential and cumulative measurements is a direct implication of the steepness of the galaxy mass function. This steepness means that at any limit, the cumulative bias is always dominated by the objects near the limit. Thus for brevity, we will focus on the analysis of large-scale differential biases in the next sections. %This is because the low luminous galaxies dominate the galaxy population.

%---------------------------------------------------------------%
%                                                               %
%                  	   	    RESULTS                             %
%                                                               %
%---------------------------------------------------------------%
\section{RESULTS}
\label{sec:results}
%                                                               %
%    					physical properties                     %
%                                                               %
\subsection{Large-scale galaxy bias dependence on physical properties}
% Figure
\begin{figure*}
    \includegraphics[width=\textwidth]{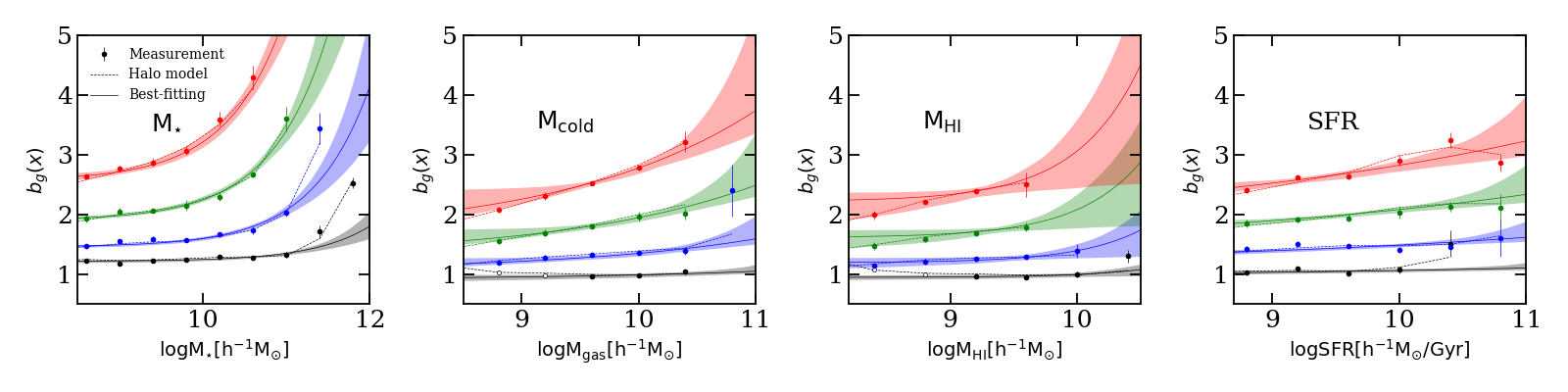}
    \includegraphics[width=\textwidth]{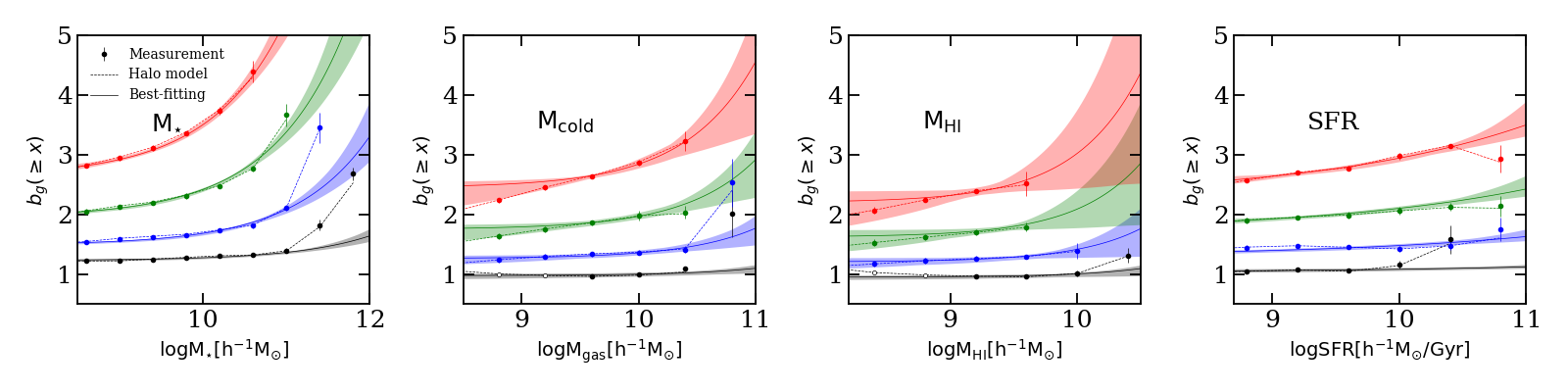}
    \caption{Large-scale galaxy bias as a function of stellar mass, cold gas mass, 
    		cold atomic gas mass and SFR at redshifts z = 0 (black), 1 (blue), 2 (green), 3(red). 
            The upper and lower panels show the differential and 
            cumulative biases respectively. 
            The filled symbols are directly measured by averaging equation~\eqref{eq:bg} on large scales.
            The error bars are estimated by propagating the shot noise of the power spectrum.
            Solid lines show the best fits of equation~\eqref{eq:fit} and
            the dashed lines show the predicted biases from the Halo model.
            Only measurements with a relative uncertainty below than 20\% are shown.
            The open circles are not used for the fitting as they lie below the resolution limit of the simulation as defined in Section \ref{sec:measurement}.
            }
    \label{fig:phys}
\end{figure*}

% Figure
\begin{figure*}
	\includegraphics[width=\textwidth]{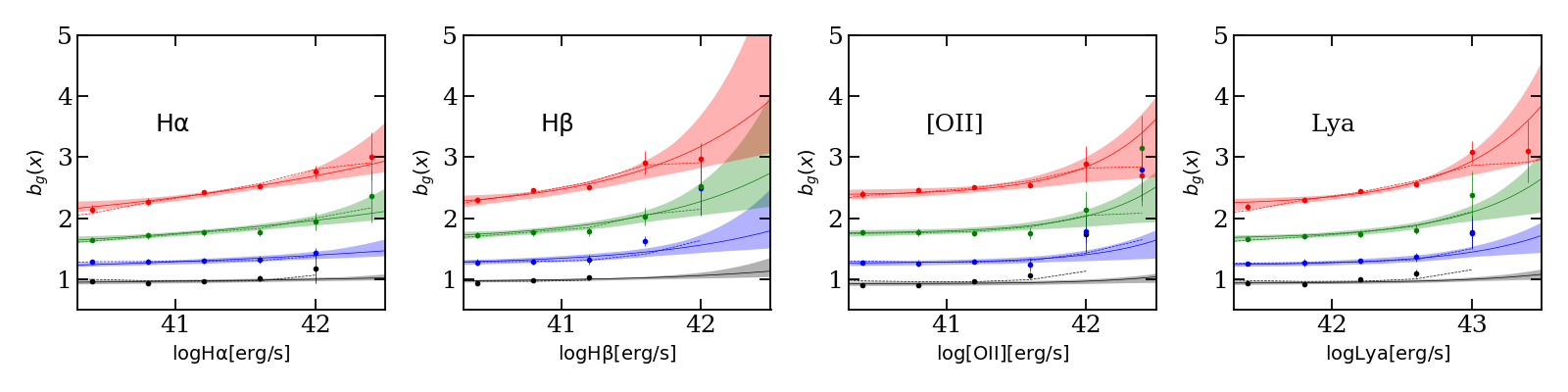}
    \includegraphics[width=\textwidth]{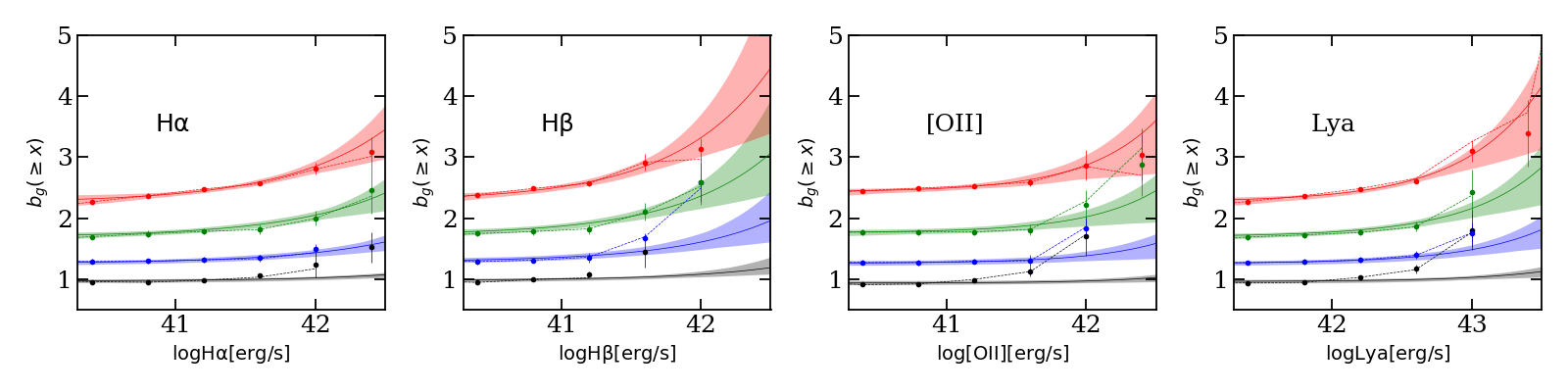}
    \caption{ Large-scale differential (top) and cumulative (bottom) galaxy bias as a function of 
              H$\alpha$, H$\beta$, [OII] and Lya line luminosities from z = 0 to z = 3
              (see caption of  Fig.~\ref{fig:phys} for details).}
    \label{fig:emib}
\end{figure*}

% Paragraph
We show the large-scale galaxy bias as a function of stellar mass, cold gas mass, cold atomic gas (\ha) mass and SFR in Fig.~\ref{fig:phys}. These properties are amongst the most critical properties in galaxy evolution and their bias will also help us understand the biases of various other observable properties. The \ha bias can be compared against measurements of the completed \ha surveys, such as HIPASS and ALFAFA \citep{barnes2001h, haynes2018arecibo}, and forecast for the next generation of extragalactic \ha surveys, such as MIGHTEE \citep{jarvis2017meerkat},  WALLABY (Koribalski et al., in preparation) and SKA \citep{yahya2015cosmological}. 

% Paragraph
One can see that the large-scale galaxy biases from the "Measurement" of selected galaxy samples are in good agreement with the "Halo model" prediction, which verifies the "Measurement" in a statistical sense and indicates a lack of assembly bias from this particular simulation in comparison to the findings in \citet{2018MNRAS.476.3631P}. The same upward trend towards the high value ends between the halo mass and galaxy physical property biases demonstrates a close correlation between these galaxy properties and halo mass as the galaxy formation models predict. Compared to the cold gas mass, \ha mass and SFR biases, the galaxy stellar mass bias shows a steeper slope, which implies that the galaxy stellar mass is fundamentally different from other galaxy properties. The cold gas and \ha gas trace the SFR, since the SFR is proportional to the mass in the molecular component which correlates with atomic hydrogen. We note the open circles in the middle panels are not used for the fitting due to the mass resolution limits of cold gas at z = 0. 

The turn-down feature of the SFR bias at high redshifts on the high star-forming end shows that dense environments have strongly reduced star formation which imposes an anti-bias effect on the distribution of star-forming galaxies at early times ($z \geq 2$). The lack of a turn-down feature in the cold gas bias plots implies the relation between cold gas and SFR is also dependent on density environments during that period of time. However, since this turn-down feature only shows in the high SFR end and the drop in numbers makes it difficult to gather enough samples for robust measurements, we look forward to seeing further verification of this trend from future simulations and observations.

%                                                               %
%    					emission lines	                        %
%                                                               %
\subsection{Large-scale galaxy bias dependence on emission lines}
% Figure
\begin{figure}
	\includegraphics[width=\columnwidth]{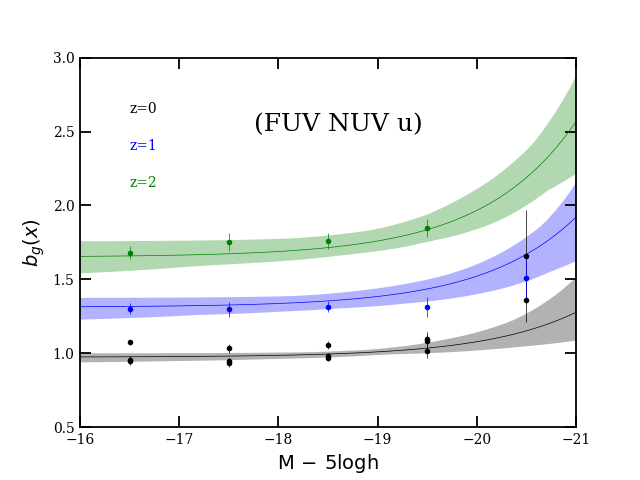}
    \includegraphics[width=\columnwidth]{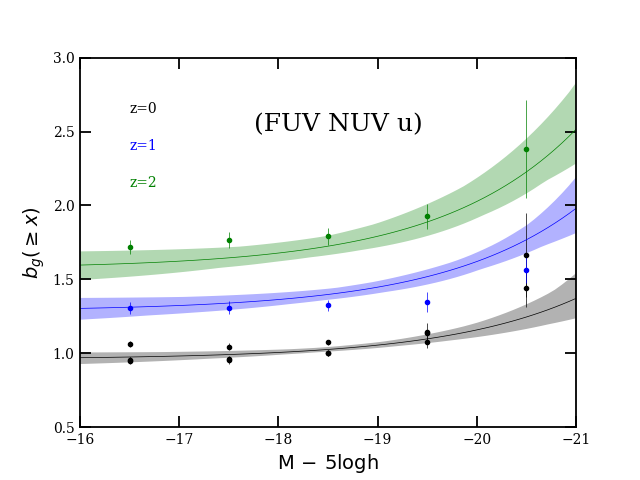}
    \caption{Large-scale differential (top) and cumulative (bottom) galaxy bias as a function of absolute magnitude
              in u band at $z = 0, 1, 2$ and FUV, NUV bands at $z = 0$ (higher redshifts not shown due to the Lyman limit).
              The M -5logh on the x-axis stands for all the magnitudes from FUV to u-band.
              The solid line shows the best-fitting formula estimated using the maximum likelihood method
              with only one set of parameters for the FUV, NUV and u bands
              (see caption of  Fig.~\ref{fig:phys} for details).}
    \label{fig:uf_uv}
\end{figure}

% Figure
\begin{figure*}
    \includegraphics[width=171mm]{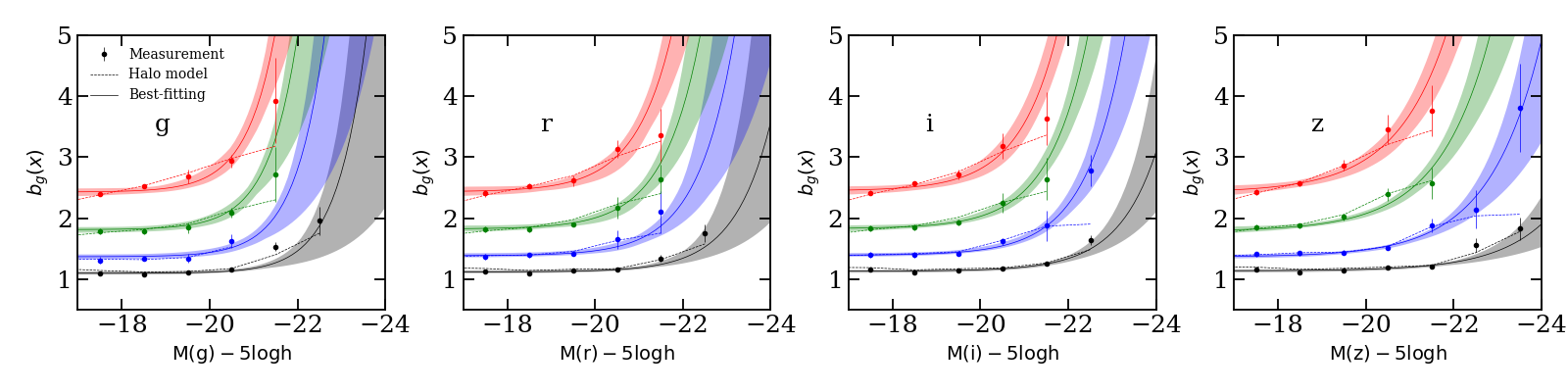}
    \includegraphics[width=171mm]{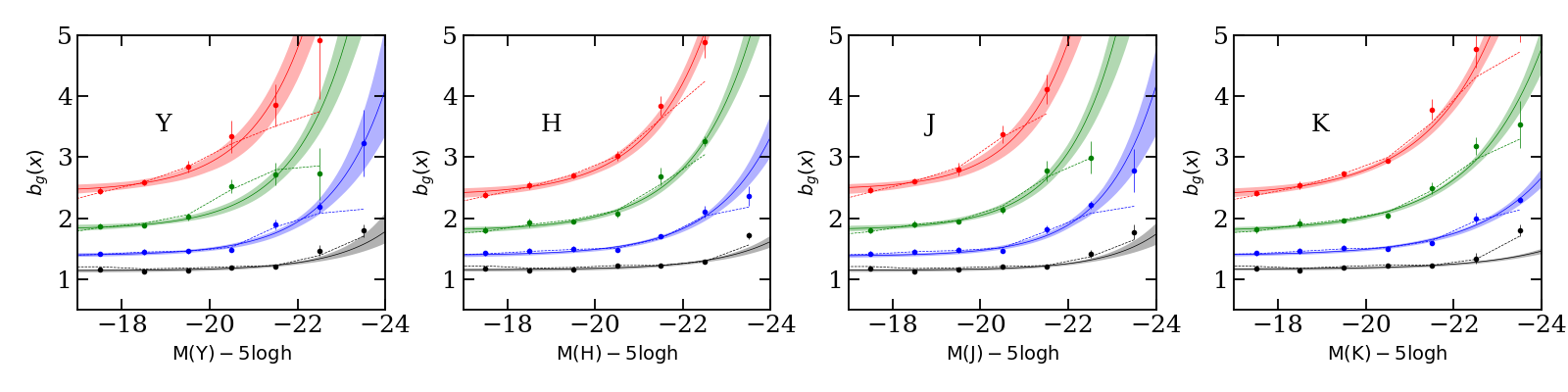}
    \includegraphics[width=171mm]{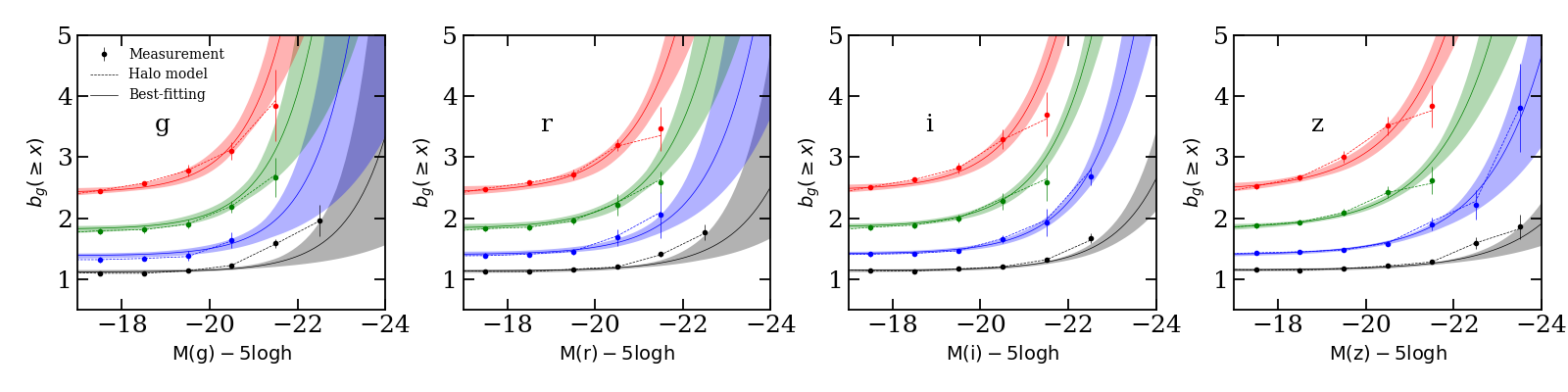}
    \includegraphics[width=171mm]{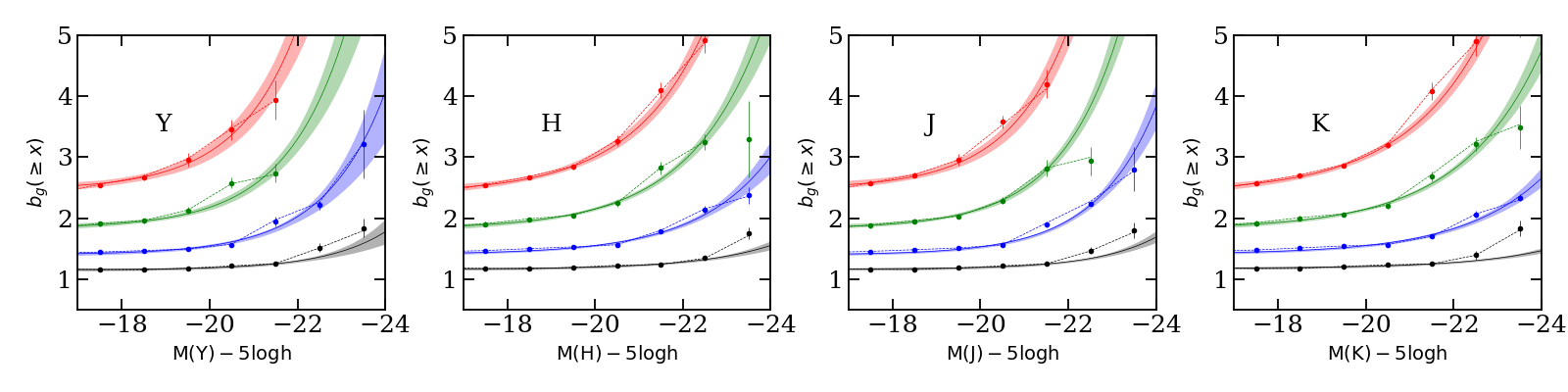}
    \caption{Large-scale galaxy bias as a function of absolute magnitude
              from optical g-band to K-band for redshifts spanning the range of $0 \leq z \leq 3$ 
              (see caption of  Fig.~\ref{fig:phys} for details).
              }
    \label{fig:continuum}
\end{figure*}

% Paragraph
We now move to the galaxy bias as a function of the luminosity for H$\alpha$, H$\beta$, Lya and [OII] lines corresponding to the wavelengths of 6563, 4861, 3727 and 1216 angstroms respectively in Fig.~\ref{fig:emib}. These emission lines are related to the galaxy selection in many current and upcoming surveys, such as UKIDSS, COSMOS, UDS, DESI, 4MOST and EUCLID \citep[e.g.][]{2012MNRAS.426..679G,levi2013desi,de20124most,laureijs2012euclid,duffy2014probing,guzzo2018measuring}.

% Paragraph
Narrowband selections allow for a clean selection of star forming galaxies based simply on the strength of an emission line sampled by the corresponding filter. For instance, the PAU Camera survey is a narrow band imaging survey that could be used to extract emission lines over the redshift range covered by GAMA \citep{stothert2018pau}. Most narrowband-selected clustering analyses conducted so-far have targeted the Lyman-$\alpha$ (Lya) emission line. The development of wide-format infrared cameras over the past decade has cleared the way for panoramic near-infrared narrowband surveys which target the H$\alpha$ nebular line.

% Paragraph
The H$\alpha$ flux of the Balmer line, created by a hydrogen atom when an electron falls from n = 3 to n = 2, is directly connected to the total hydrogen-ionizing radiation from massive stars, making it a reliable tracer of star formation. The H$\alpha$ emission line  has been one of the primary diagnostics used to estimate the SFRs of galaxies in the local universe \citep{1983ApJ...272...54K}, although the measurements are complicated by dust absorption of Lyman-continuum photons within individual \htwo regions, dust attenuation in the general interstellar medium of galaxy and uncertainties in the shape of the initial mass function \citep{1998ARA&A..36..189K}. 

% Paragraph
Above z $\sim$ 0.4, H$\alpha$ becomes inaccessible to ground-based optical spectrographs, the higher-order Balmer lines such as H$\beta$ offer a promising alternative. H$\beta$, like all the Balmer lines, inherits the same strength and weaknesses of H$\alpha$:  it is equally sensitive to variations in the IMF and the absorption of Lyman-continuum photons within star forming regions. Furthermore, H$\beta$ suffers more interstellar dust attenuation and is more sensitive to underlying stellar absorption. Despite these uncertainties, H$\beta$ may be a superior SFR diagnostic than the more commonly used [OII] nebular emission line \citep{2006ApJ...642..775M}.

% Paragraph
The [OII] emission line has also been used widely as a qualitative and quantitative tracer of star formation in galaxies in redshift ranges where the H$\alpha$ emission line moves into the near-infrared \citep{2005MNRAS.362.1143M}. However, SFRs based on [OII] are still subject to considerable systematic uncertainties due to variations in dust reddening, chemical abundance, and ionization among star-forming galaxies \citep{2006ApJ...642..775M}.

% Paragraph
One of the most promising ways of detecting very high redshift (z $\gtrsim$ 5) star-forming galaxies is via narrow-band imaging surveys targeting Lya. Lya emission originates from reprocessed ionizing photons of massive stars. The ionizing photons ionize the neutral hydrogen atoms in the interstellar medium (ISM). As a consequence of radiative transfer, H$\alpha$ photons can also be transfered into Lya, which makes the Lya emission line stronger than others as shown in the parameter $c$ for the emission lines. 

% Paragraph
Fig.~\ref{fig:emib} shows that generally all these emission lines can be treated as fine tracers of the star formation rate of galaxy (in our semi-analytic model). The turn-down of the Lya, H$\beta$ and [OII] bias at the highest luminosities and z = 3 reflects the same feature seen in Fig. 6 for SFR.  The H$\alpha$ line does not show an obvious turn-down feature as the others do, which indicates that the H$\alpha$ line could be a better tracer of cold gas than the other lines. However, on the luminous ends, the discrepancy between the "Measurement" and "Halo model" suggests that the systematic uncertainties play a key role in the measurements as explained above. The galaxy assembly bias may also contribute to this discrepancy. However, it would be very difficult to detect without including a third property such as halo formation time in addition to halo mass. This would be an interesting thing to look at in future work. A similar discrepancy also arises for the measurements in the continuum bands as seen in the next section.

%                                                               %
%    					continuous bands                        %
%                                                               %
\subsection{Large-scale galaxy bias dependence on continuum bands}
% Figure
\begin{figure*}
    \includegraphics[width=171mm]{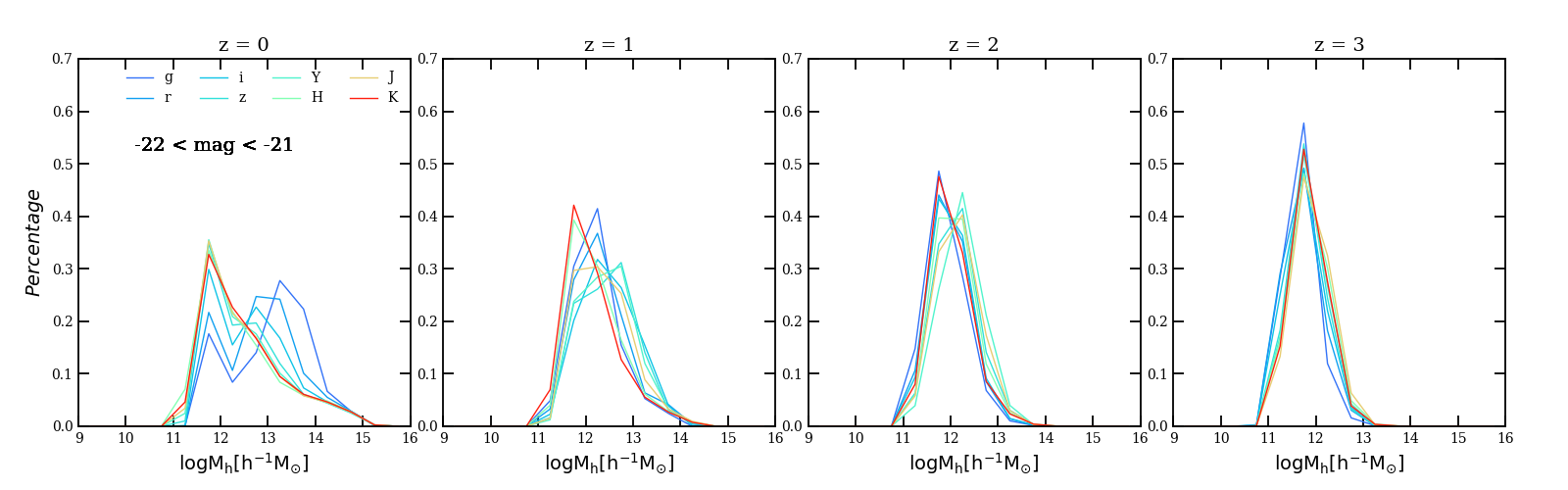} 
    \includegraphics[width=171mm]{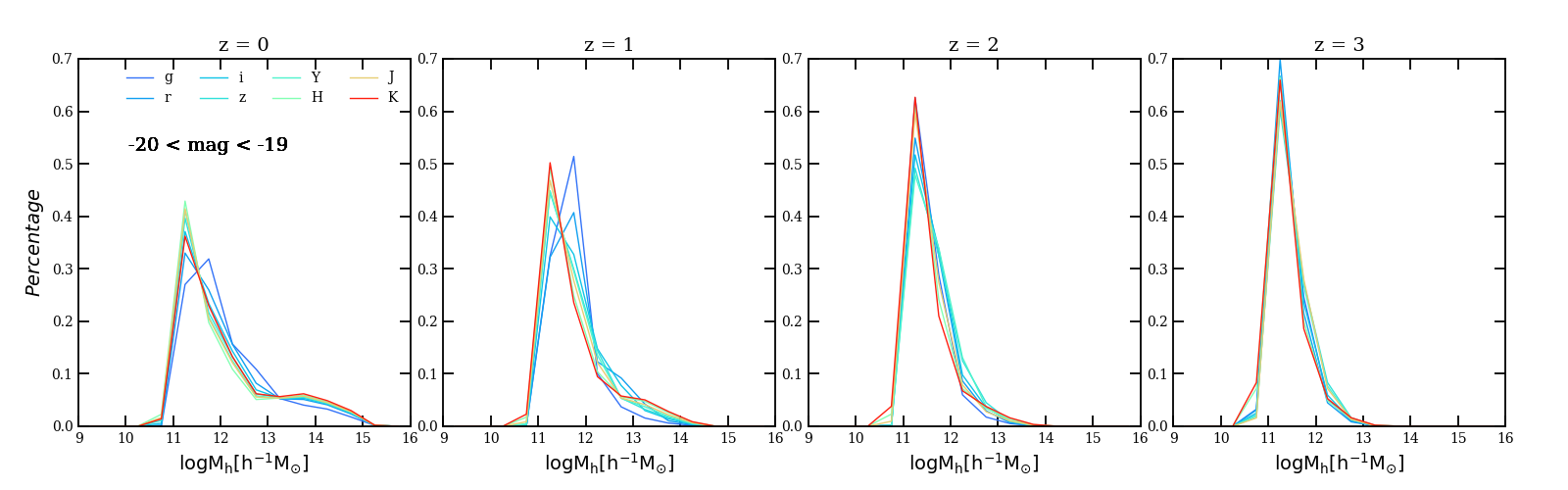} 
    \caption{The host halo mass distribution at redshifts 0, 1, 2, 3 from left panel to right panel.
    		 The distribution of host halo mass are plotted for two different ranges of 
             absolute magnitude as indicated in left panels.
             The lines are color-coded by filters from g-band to K-band.}
    \label{fig:halohis}
\end{figure*}

% Paragraph

% Paragraph
In Fig.~\ref{fig:uf_uv} and~\ref{fig:continuum}, we show the galaxy bias as a function of observer-frame absolute magnitude for UV bands and optical/IR bands. These filters are widely used in the ongoing and upcoming LSS surveys such as BOSS, eBOSS, DES and LSST surveys \citep{dawson2012baryon, dawson2016sdss, dark2005dark, tyson2002large}.

% Paragraph
Shorter rest-frame wavelengths than 91nm hardly escape the galaxies and are virtually irrelevant for LSS studies, thus we only show the FUV, NUV bands biases at $z = 0$ and u-band bias at $0 \leq z \leq 2$. From optical g-band to K-band, we are able to cover the redshifts spanning the range $0 \leq z \leq 3$.

% Paragraph
In Fig.~\ref{fig:uf_uv}, one can see that the steep slopes are similar to that of emission line biases, which shows a strong correlation between UV continuum and emission lines. In Fig.~\ref{fig:continuum}, all the measured biases from optical g-band to K-band show a self-similar dependence on the magnitude. The physical processes behind UV and IR bands emission on the rest frame are quite different. Like the emission lines, the UV bands are sensitive to the instantaneous and unobscured star formation, as they are driven by massive young stars. Conversely the IR-band is more sensitive to the stellar mass of the galaxy, especially on the luminous end, since the massive red galaxies have already consumed most of their cold gas and live in over dense environments where only few star forming galaxies reside. 

% Paragraph
In most of the visible bands in the observer frame, contribution from the massive red galaxies and star forming galaxies are mixed, thus bias similarities emerge across those bands. Nevertheless, one can still see that the slopes of biases are decreasing when you look at them from high-energy filter (UV) to low-energy filter (z-band).

% Paragraph
\begin{figure*}
\centering
  \begin{subfigure}{\columnwidth}
    \centering
    \includegraphics[width=\columnwidth]{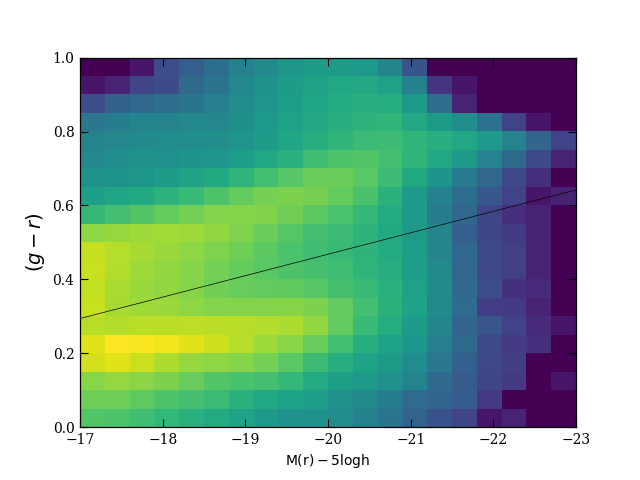}
    \label{fig:sub1}
  \end{subfigure}%
  \begin{subfigure}{\columnwidth}
    \centering
    \includegraphics[width=\columnwidth]{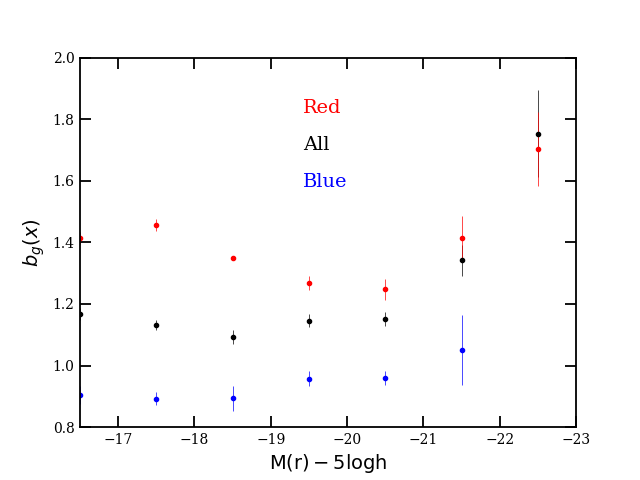}
    \label{fig:sub2}
  \end{subfigure}
  \caption{Left: (g-r) color magnitude diagram at $z = 0$.
              The tilted line defined by (g-r) $=-0.058[$M(r)-5$\log$h$]-0.692$ divides the red and blue populations.
              Right: The galaxy bias as a function of r-band magnitude at $z = 0$ 
              in blue clouds and red sequence shown in corresponding colors.}
  \label{fig:gr}
\end{figure*}

% Paragraph
We can understand the similar galaxy bias dependence on continuum bands by investigating the host halo distribution. In Fig.~\ref{fig:halohis} we show the host halo mass distribution in magnitude bins for redshifts z = 0, 1, 2, 3. Comparing the colored lines, we see that the effects of filters on the host halo distribution are quite small in this particular SAM, except for the brightest objects at z = 0. We find that many of the brightest objects at z = 0 reside in very massive halos. A plausible cause is the high merger efficiency in the current universe. High merger efficiency generally means high star formation rate; therefore, the luminous galaxies at $z = 0$ with high star formation rate are clustered more than the galaxies with low star formation rate so that the slopes of the biases decrease from UV to z-band. When compared to the turn-down feature at high redshift of the SFR bias, this indicates that the large-scale density environments have changing effects on the large-scale clustering of star-forming galaxies as the universe evolves.

To expand our results, we study the bias dependence on galaxy colors in Fig.~\ref{fig:gr}. We show the color magnitude diagram and differential bias as a function of r-band magnitude at $z = 0$ in the left and right panels. Based on the clear bimodal color distribution, we split the galaxies into red and blue populations using the black line shown in the left panel and produce the biases of the red and blue population respectively. As expected, the distribution of red galaxies is much more clustered against the dark matter background than that of the blue galaxies. This has been well understood within the halo model framework  \citep{2009MNRAS.392.1080S}. For the fainter red galaxies, the increasing bias implies that they are mostly the low mass satellites in large host halos (also indicated in Fig.~\ref{fig:hvm}). The trend of bias dependence on galaxy color is consistent with the findings of \citet{zehavi2011galaxy} who investigated the clustering strength with the correlation length. Although there is a significant difference in bias between star forming and passive galaxies caused by the types of halos in which they reside, in a magnitude limited survey the contributions from these two populations is mixed together and hence the bias shows only a dependence on wavelength for the most luminous objects at low redshifts. %This fully explains the similarities we see in bias across the g-K continuum bands.

\subsection{Large-scale galaxy bias as a function of wavelength}
\begin{figure}
    \includegraphics[width=\columnwidth]{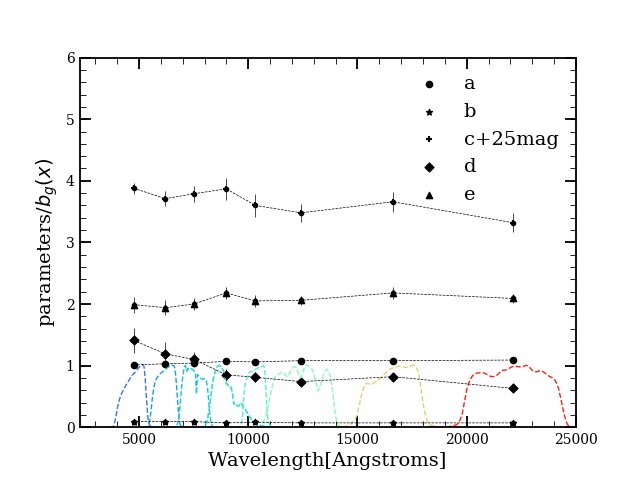}
    \includegraphics[width=\columnwidth]{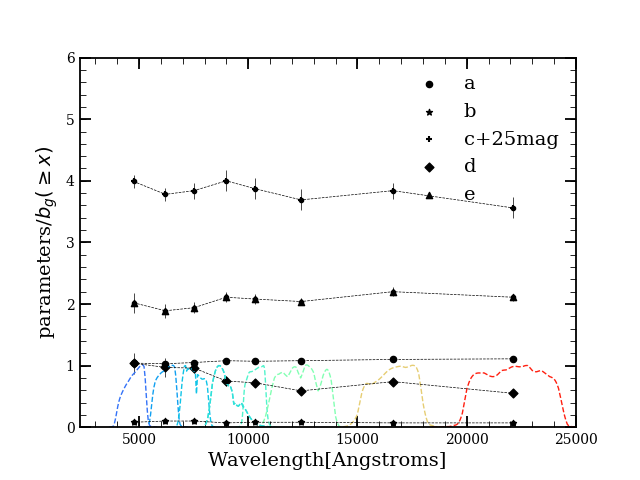}
    \caption{The best-fitting parameters for equation~\eqref{eq:fit} as a function of wavelength
    		from g-band to K-band. 
            The upper and lower panels show the parameters for fitting the differential and 
            cumulative galaxy biases respectively. 
            The parameter c is represented by adding 25 mag for convenience
            The dashed lines are color-coded by the observational filters 
            from g-band to K-band. 
            }
    \label{fig:parameters}
\end{figure}

We now look at how our bias model depends on wavelengths. Fig.~\ref{fig:parameters} shows the five parameters for our fits of equation~\eqref{eq:fit} from g-band to K-band plotted at the effective wavelength of each filter. The decrease of the parameter $d$ (defined in Section \ref{sec:model}) towards higher wavelengths means that the slopes of biases get steeper when we look at them from low-energy filter (z-band) to high-energy filter (UV), as explained via Fig. \ref{fig:halohis} and \ref{fig:gr} in the last section.  
% Figure
% Figure
\begin{figure}
	\includegraphics[width=\columnwidth]{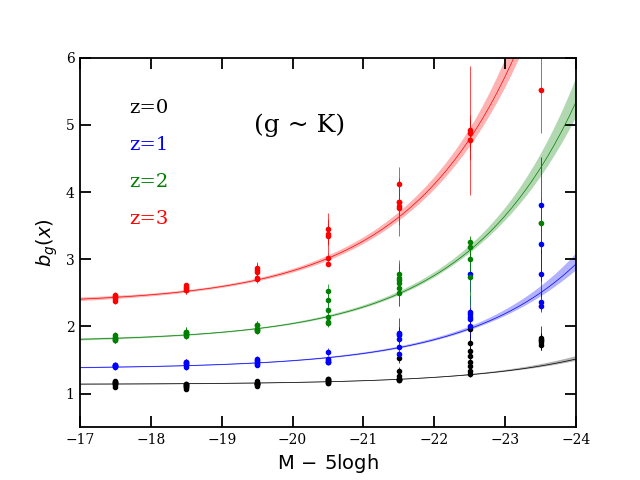}
    \includegraphics[width=\columnwidth]{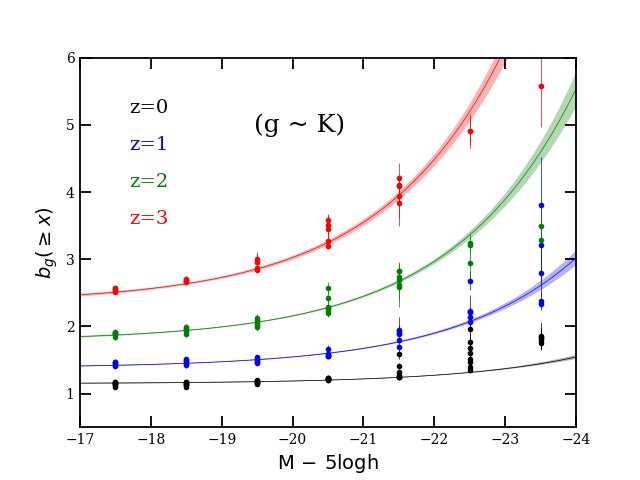}
    \caption{Large-scale galaxy bias as a function of absolute magnitude 
              from g-band to K-band for redshifts spanning the range $0 \leq z \leq 3$.
              The M -5logh on the x-axis stands for all the magnitudes from g-band to K-band.
              The solid line shows the fit of equation~\eqref{eq:fit}
              from g-band to K-band with only one set of parameters
              (see caption of  Fig.~\ref{fig:phys} for details).}
    \label{fig:uf}
\end{figure}

The other parameters appear to be constant (within their statistical uncertainties) over the considered wavelength range. Thus, we can attempt to model all wavelengths using a single set of parameters (allowing a small systematic error on $d$). To constrain these parameters, we take all the measurements at all wavelengths as the input of equation \eqref{eq:sbga}. The resulting parameters are provided at the bottom of Table \ref{params_db} and \ref{params_cb}. This universal 5-parameter fit for all wavelengths (and redshifts) is shown in Figure 13. The slight discrepancy between the fits (lines) and the simulated data (dots) for the brightest galaxies at z = 0 is due to the assumption of a universal d-parameter, which is not strictly correct in a statistical sense. Of course, better fits can always be obtained by using the wavelength-dependent parameters also given in Table \ref{params_db} and \ref{params_cb}.

\section{DISCUSSIONS}
\label{sec:discussions}
\subsection{Bias forecasts in observer frame quantities}
\label{sec:Survey}
% Paragraph
For the convenience of estimating the biases in LSS surveys with fixed redshift ranges, we convert the absolute magnitude and luminosity into the apparent magnitude and flux. We provide the apparent magnitude limited bias as a function of apparent magnitude and redshift for the broad-band filters by
\begin{equation}
	b_g(\leq m, z_1,z_2) = \frac{\int_{z_1}^{z_2}\int_{-\infty}^{M(m,z)} 
    					 b_g(M',z)\phi(M',z)dM'\frac{dV}{dz}dz}
                         {\int_{z_1}^{z_2}\int_{-\infty}^{M(m,z)}\phi(M',z)dM'\frac{dV}{dz}dz},
    \label{eq:bmz}
\end{equation}
where m is the apparent magnitude and $b_g(M',z)$ is the large-scale differential bias as a function of absolute magnitude and redshift.

% Paragraph
Likewise, we provide the flux limited bias as a function of flux and redshift for the emission lines by
\begin{equation}
%\begin{aligned}
	b_g(\geq F, z_1,z_2) = \frac{\int_{z_1}^{z_2}\int_{L(F,z)}^{\infty} 
    					b_g(L',z)\phi(L',z)dL'\frac{dV}{dz}dz}
                        {\int_{z_1}^{z_2}\int_{L(F,z)}^{\infty}\phi(L',z)dL'\frac{dV}{dz}dz},
    \label{eq:bfz}
%\end{aligned}
\end{equation}
where F is the measured flux and $b_g(L',z)$ is the large-scale differential bias as a function of luminosity and redshift. 

% Paragraph
The conversions between $m, F$ and $M, L$ are given via %$m = M + 5log(hD_L(Mpc)) + 25, F = \frac{L}{4\pi D_L^2}$
\begin{equation} 
	M(m,z) = m - 5\log \left(\frac{d_L(z)}{Mpc}\right) - 25 + 2.5\log(1+z),
\end{equation}
\begin{equation} 
	L(F,z) = F \times 4\pi d_L^2(z),
\end{equation}
where $d_L(z) = \frac{(1+z)}{H_0} \int_0^z \frac{dz^\prime}{\sqrt{\Omega_\Lambda +
\Omega_m(1+z^\prime)^3}}$ is the luminosity distance to the galaxy in unit of Mpc. Note that the conversion between \ha mass and the flux of \ha emission line is given via $M_{\rm HI} = 2.356 \times 10^5 d^2_L(z)(1+z)^{-1} S $, where the $M_{\rm HI}$ is in solar mass and the $S$ is an integrated flux in unit of Jy km/s. The $(1+z)$ factor is needed since the integrated flux $S$ is expressed in units of Jy km/s rather than Jy Hz \citep[see Appendix A in][]{obreschkow2009heuristic}. 

\subsection{Comparison with existing surveys and forecasts}
% Paragraph
In this section, we compare the predictions obtained using the equations of Section \ref{sec:Survey} with existing surveys and forecasts. The results are listed in Table \ref{tab:survey}. 

% Paragraph
HIPASS was a blind survey of neutral atomic hydrogen (\ha), which covered 71\% of the sky and identified more than 5000 galaxies below $z \sim 0.02$. \cite{basilakos2007large} measure the overall linear bias of 1619 \ha galaxies more massive than $1.89 \times 10^9 h^{-1}\rm M_\odot$ (calibrated by the Planck 2015 cosmology) using a correlation function analysis. The ALFALFA survey is a census of galaxies in the local universe, out to $z \sim 0.06$, with much better resolution \citep{haynes2018arecibo}. \cite{martin2012clustering} used the $\alpha$.40 sample of ALFAFA containing the results of the 40\% survey to investigate the bias for \ha-selected objects. They found the sample became unbiased (i.e. $b_g = 1$) on large scales. The latest forecast of \ha bias from the \textsc{GALFORM} in \cite{baugh2018galaxy} (hereafter B18) is an intensity mapping prediction that shows the evolution of \ha bias up to $z = 3$ including all the galaxies within a halo. The B18 forecasts are roughly estimated from the Fig. 11 in that paper. We predict the \ha bias of galaxies more massive than $M_{\rm HI}=10^7 h^{-1}\rm M_\odot$ for comparison with the B18.

% Paragraph
EUCLID \citep{laureijs2011euclid} is a space-based survey mission designed to understand the origin of the Universe's accelerating expansion using two independent primary cosmological probes: Weak gravitational Lensing (WL) and Baryonic Acoustic Oscillations (BAO). The BAO are determined from a spectroscopic survey predominantly detecting H$\alpha$ emission line galaxies. \cite{amendola2017constraints} forecast the errors on the H$\alpha$ galaxy bias based on  a simple power-law model and the polynomial model proposed by \cite{cole20052df}. 

% Paragraph
The 6dFGS \citep{jones20096df} is a near-infrared selected ($J H K$ ) redshift survey of 125 000 galaxies across four-fifths of the southern sky. \cite{beutler20126df} measured the K-band bias from 6dFGS by exploiting the angular dependence of redshift-space distortions in the 2D correlation function at effective redshift 0.067. Note that we compute the K-band cumulative bias using two sets of best-fitting parameters from K-band in Fig. \ref{fig:continuum} and g$\sim$K band in Fig. \ref{fig:uf} separated by a slash in Table~\ref{tab:survey}.

% Paragraph
Comparing the results from our 5-parameter model with existing surveys (HIPASS, ALFAFA, 6dFGS) and forecasts (EUCLID), we find that our model is in good agreement with these references within $1\sigma$ statistical uncertainties especially when accounting for the fact that we all use different background cosmologies and methodologies. However, our model does predict a higher bias beyond the $1\sigma$ uncertainty when compared to the B18 forecast. We elaborate on some limitations in the next section.
% Table
\begin{table*}
	\centering
	\caption{The best-fitting parameters of our 5-parameter model for the large-scale \textit{differential} bias for different observational bands, emission lines, physical properties of galaxies and halo mass as well.}
	\begin{tabular}{llccccr} % four columns, alignment for each
		\hline
        \hline
		Halo mass & x & a & b & c & d &e\\
        $\mathrm{M_{h}}$ & $\mathrm{logM_{h}[h^{-1}M_{\odot}]}$ & 0.609$\pm$0.024 & 0.035$\pm$0.007 & 10.893$\pm$0.286 & 1.057$\pm$0.066 & 2.172$\pm$0.086 \\

		\hline
        Physical property & x & a & b & c & d & e \\
$\mathrm{M_{\star}}$ & $\mathrm{logM_{\star}[h^{-1}M_{\odot}]}$ & 1.146$\pm$0.015 & 0.069$\pm$0.008 & 10.527$\pm$0.065 & 1.435$\pm$0.145 & 2.172$\pm$0.078 \\
$\mathrm{M_{gas}}$ & $\mathrm{logM_{gas}[h^{-1}M_{\odot}]}$ & 0.86$\pm$0.034 & 0.076$\pm$0.023 & 10.483$\pm$0.605 & 0.925$\pm$0.431 & 1.946$\pm$0.119 \\
$\mathrm{M_{HI}}$ & $\mathrm{logM_{HI}[h^{-1}M_{\odot}]}$ & 0.872$\pm$0.033 & 0.071$\pm$0.026 & 10.319$\pm$0.596 & 1.206$\pm$0.571 & 2.004$\pm$0.167 \\
SFR & $\mathrm{logSFR[h^{-1}M_{\odot}/Gyr]}$ & 0.868$\pm$0.031 & 0.129$\pm$0.034 & 11.068$\pm$0.781 & 0.537$\pm$0.313 & 1.643$\pm$0.082 \\

		\hline
        Emission line & x & a & b & c & d & e \\
$\mathrm{H{\alpha}}$ & $\mathrm{logH{\alpha}[erg/s]}$ & 0.831$\pm$0.032 & 0.097$\pm$0.023 & 42.49$\pm$0.571 & 0.639$\pm$0.3 & 1.736$\pm$0.108 \\
$\mathrm{H{\beta}}$ & $\mathrm{logH{\beta}[erg/s]}$ & 0.829$\pm$0.032 & 0.12$\pm$0.028 & 42.147$\pm$0.252 & 1.038$\pm$0.468 & 1.676$\pm$0.129 \\
$\mathrm{[OII]}$ & $\mathrm{log[OII][erg/s]}$ & 0.786$\pm$0.036 & 0.134$\pm$0.026 & 42.918$\pm$0.412 & 1.407$\pm$0.558 & 1.748$\pm$0.113 \\
Lya & $\mathrm{logLya[erg/s]}$ & 0.811$\pm$0.032 & 0.117$\pm$0.022 & 43.417$\pm$0.139 & 1.279$\pm$0.522 & 1.726$\pm$0.115 \\

        \hline
        Bands & x & a & b & c & d & e \\
        (FUV NUV u) & M - 5logh & 0.629$\pm$0.187 & 0.338$\pm$0.184 & -21.201$\pm$0.291 & 1.012$\pm$0.235 & 0.997$\pm$0.305 \\

        g & $\mathrm{M(g)-5logh}$ & 1.009$\pm$0.025 & 0.09$\pm$0.017 & -21.118$\pm$0.09 & 1.41$\pm$0.198 & 1.985$\pm$0.133 \\
r & $\mathrm{M(r)-5logh}$ & 1.026$\pm$0.024 & 0.095$\pm$0.017 & -21.292$\pm$0.116 & 1.194$\pm$0.201 & 1.945$\pm$0.118 \\
i & $\mathrm{M(i)-5logh}$ & 1.045$\pm$0.02 & 0.087$\pm$0.013 & -21.214$\pm$0.132 & 1.102$\pm$0.115 & 1.999$\pm$0.099 \\
z & $\mathrm{M(z)-5logh}$ & 1.073$\pm$0.017 & 0.066$\pm$0.01 & -21.134$\pm$0.177 & 0.852$\pm$0.095 & 2.177$\pm$0.097 \\

        Y & $\mathrm{M(Y)-5logh}$ & 1.059$\pm$0.018 & 0.081$\pm$0.012 & -21.403$\pm$0.182 & 0.808$\pm$0.079 & 2.047$\pm$0.096 \\
H & $\mathrm{M(H)-5logh}$ & 1.077$\pm$0.016 & 0.074$\pm$0.009 & -21.518$\pm$0.148 & 0.739$\pm$0.052 & 2.064$\pm$0.072 \\
J & $\mathrm{M(J)-5logh}$ & 1.077$\pm$0.017 & 0.066$\pm$0.01 & -21.339$\pm$0.16 & 0.819$\pm$0.065 & 2.183$\pm$0.095 \\
K & $\mathrm{M(K)-5logh}$ & 1.094$\pm$0.015 & 0.069$\pm$0.007 & -21.681$\pm$0.149 & 0.63$\pm$0.036 & 2.095$\pm$0.069 \\

        $\mathrm{(g \sim K)}$ & M - 5logh & 1.062$\pm$0.007 & 0.075$\pm$0.004 & -21.462$\pm$0.067 & 0.645$\pm$0.018 & 2.044$\pm$0.033 \\

	\end{tabular}
    \label{params_db}
\end{table*}

% Table
\begin{table*}
	\centering
	\caption{The best-fitting parameters of our 5-parameter model for the large-scale \textit{cumulative} galaxy bias for different observational bands, emission lines and physical properties of galaxies.}
	\begin{tabular}{llccccr} % four columns, alignment for each
		\hline
		\hline
		Halo mass & x & a & b & c & d &e\\
        $\mathrm{M_{h}}$ & $\mathrm{logM_{h}[h^{-1}M_{\odot}]}$ & 0.603$\pm$0.025 & 0.066$\pm$0.007 & 11.305$\pm$0.147 & 1.142$\pm$0.055 & 1.99$\pm$0.055 \\

		\hline
        Physical property & x & a & b & c & d & e \\
$\mathrm{M_{\star}}$ & $\mathrm{logM_{\star}[h^{-1}M_{\odot}]}$ & 1.157$\pm$0.013 & 0.075$\pm$0.006 & 10.488$\pm$0.064 & 1.126$\pm$0.098 & 2.149$\pm$0.054 \\
$\mathrm{M_{gas}}$ & $\mathrm{logM_{gas}[h^{-1}M_{\odot}]}$ & 0.883$\pm$0.032 & 0.082$\pm$0.023 & 10.756$\pm$0.452 & 1.055$\pm$0.524 & 2.015$\pm$0.116 \\
$\mathrm{M_{HI}}$ & $\mathrm{logM_{HI}[h^{-1}M_{\odot}]}$ & 0.864$\pm$0.03 & 0.083$\pm$0.027 & 10.516$\pm$0.661 & 1.263$\pm$0.577 & 1.94$\pm$0.141 \\
SFR & $\mathrm{logSFR[h^{-1}M_{\odot}/Gyr]}$ & 0.942$\pm$0.025 & 0.098$\pm$0.021 & 11.161$\pm$0.484 & 0.613$\pm$0.259 & 1.886$\pm$0.078 \\

		\hline
        Emission line & x & a & b & c & d & e \\
$\mathrm{H{\alpha}}$ & $\mathrm{logH{\alpha}[erg/s]}$ & 0.844$\pm$0.031 & 0.116$\pm$0.02 & 42.623$\pm$0.132 & 1.186$\pm$0.387 & 1.765$\pm$0.106 \\
$\mathrm{H{\beta}}$ & $\mathrm{logH{\beta}[erg/s]}$ & 0.837$\pm$0.036 & 0.135$\pm$0.027 & 42.2$\pm$0.119 & 1.409$\pm$0.395 & 1.681$\pm$0.117 \\
$\mathrm{[OII]}$ & $\mathrm{log[OII][erg/s]}$ & 0.816$\pm$0.03 & 0.118$\pm$0.022 & 42.993$\pm$0.407 & 1.38$\pm$0.535 & 1.855$\pm$0.109 \\
Lya & $\mathrm{logLya[erg/s]}$ & 0.834$\pm$0.031 & 0.119$\pm$0.021 & 43.396$\pm$0.102 & 1.526$\pm$0.4 & 1.754$\pm$0.11 \\

        \hline
        Bands & x & a & b & c & d & e \\
        (FUV NUV u) & M - 5logh & 0.477$\pm$0.202 & 0.479$\pm$0.205 & -21.209$\pm$0.422 & 0.73$\pm$0.17 & 0.744$\pm$0.194 \\

        g & $\mathrm{M(g)-5logh}$ & 1.036$\pm$0.028 & 0.083$\pm$0.02 & -21.009$\pm$0.115 & 1.035$\pm$0.174 & 2.022$\pm$0.157 \\
r & $\mathrm{M(r)-5logh}$ & 1.035$\pm$0.026 & 0.101$\pm$0.017 & -21.215$\pm$0.109 & 0.968$\pm$0.146 & 1.893$\pm$0.114 \\
i & $\mathrm{M(i)-5logh}$ & 1.046$\pm$0.02 & 0.096$\pm$0.013 & -21.165$\pm$0.125 & 0.965$\pm$0.083 & 1.939$\pm$0.088 \\
z & $\mathrm{M(z)-5logh}$ & 1.078$\pm$0.017 & 0.073$\pm$0.01 & -21.003$\pm$0.17 & 0.751$\pm$0.078 & 2.11$\pm$0.081 \\

        Y & $\mathrm{M(Y)-5logh}$ & 1.074$\pm$0.017 & 0.078$\pm$0.01 & -21.133$\pm$0.173 & 0.718$\pm$0.059 & 2.077$\pm$0.085 \\
H & $\mathrm{M(H)-5logh}$ & 1.084$\pm$0.014 & 0.078$\pm$0.008 & -21.306$\pm$0.17 & 0.593$\pm$0.038 & 2.037$\pm$0.054 \\
J & $\mathrm{M(J)-5logh}$ & 1.097$\pm$0.013 & 0.066$\pm$0.007 & -21.163$\pm$0.128 & 0.738$\pm$0.039 & 2.203$\pm$0.066 \\
K & $\mathrm{M(K)-5logh}$ & 1.106$\pm$0.014 & 0.071$\pm$0.007 & -21.443$\pm$0.167 & 0.552$\pm$0.03 & 2.11$\pm$0.058 \\

        $\mathrm{(g \sim K)}$ & M - 5logh & 1.073$\pm$0.006 & 0.075$\pm$0.003 & -21.114$\pm$0.076 & 0.574$\pm$0.015 & 2.04$\pm$0.026 \\

	\end{tabular}
    \label{params_cb}
\end{table*}

% Table
\begin{table*}
	\centering
	\caption{The comparison of our 5-parameter model with existing surveys and forecasts. The uncertainties of our prediction is calculated through error propagation of the five parameters.  Note that the B18 is based on an intensity mapping prediction without mass cut on \ha samples. The $b_g$(Survey/Forecast) and $b_g(x,z)$ stand for the large-scale bias estimated in the existing survey/forecast and predicted from our model respectively.}
	\begin{tabular}{llccr} % four columns, alignment for each
		\hline
        \hline
		Survey/Forecast & Selection criteria       &z & $b_g$(Survey/Forecast)  & $b_g(x,z)$\\
        \hline
        HIPASS & $\rm M_{\rm HI}  > 1.89 \times 10^{9} h^{-1}\rm M_\odot$ &0 & 0.94$\pm$0.15 & 0.96$\pm$0.05 \\
        B18 & -  &[0, 1, 2, 3] & [0.63, 0.8, 1, 1.26] & [0.95$\pm$0.04, 1.19$\pm$0.11, 1.58$\pm$0.26, 2.1$\pm$0.5] \\
        EUCLID & $\mathrm{F(H\alpha) > 3 \times 10^{-16} erg \; cm^{-2} s^{-1}}$ &1 & 1.36$\pm$0.03 & 1.49$\pm$0.14 \\
        6dFGS & m(K) < 12.75 mag &0.067 & 1.48$\pm$0.27 & 1.34$\pm$0.03/1.36$\pm$0.02 \\
        
	\end{tabular}
    \label{tab:survey}
\end{table*}

\subsection{Limitations of our method}
\label{sec:limit}
% Paragraph
As seen, our 5-parameter model reproduces the bias reasonably well compared with existing surveys and forecasts, although there are some caveats which one needs to be aware of.

% Paragraph
First, we only fit the bias measurements using certain lower limits and extrapolate below this. If we expect the lowest \ha mass galaxies to have a bias less than unity, then the inclusion of galaxies from $M_{\rm HI}=10^7 h^{-1}\rm M_\odot$ to $M_{\rm HI}=10^9 h^{-1}\rm M_\odot$ (i.e. roughly matching the B18 selection) could reduce the overall galaxy bias for a mass limited survey and bring our predictions closer to those of B18. In general, we do not expect our model to perform well in surveys with fainter limits than the left ends shown in the bias plots.

% Paragraph
In addition, our models cannot capture the upturn at low redshifts very well. Including this sharp increase at z = 0 will make our 6dFGS prediction slightly higher (i.e. closer to the 6dFGS survey) and this is an obvious place for improvement in future work. Similarly, our model does not reproduce the turn-down feature at high SFR and high redshifts. We expect more evidence from the upcoming surveys to verify this. % As for the turn-down features of high redshifts manifested in the SFR and most of emission lines bias plots

% Paragraph
A final caveat is that our model is based on \textsc{GALFORM} galaxies tuned with a Plank cosmology simulation. Therefore those fitting parameters should be dependent on this particular SAM and the cosmology used in there. Although using a SAM has many advantages as mentioned in Section~\ref{sec:introduction}, we note that SAMs are not perfect. For instance, SAMs have a large parameter set such that the degeneracies between these parameters are not clear,  therefore there are systematics in the underlying data sets used for fixing the SAM parameters when comparing derived quantities such as the stellar mass function. Thus a comparison of how the fitting values of our 5-parameter model change when the same procedure is applied to a hydro simulation such as the EAGLE would be interesting. Of course, using a range of SAMs and hydro simulations calibrated to fit both the same data sets and different ones should provide a better insight. 

%---------------------------------------------------------------%
%                                                               %
%           DISCUSSIONS AND CONCLUTIONS                         %
%                                                               %
%---------------------------------------------------------------%
\section{CONCLUSIONS}
\label{sec:conclusions}
% Paragraph
In conclusion, we use the \textsc{GALFORM} galaxy formation model to predict the large-scale galaxy bias as a function of redshift and magnitude threshold for broadband continuum emission from the far infrared to ultra-violet, as well as for prominent emission lines, such as the H$\alpha$, \ha lines and and intrinsic physical galaxy properties. We provide the fitting formula $b_g(x,z)=a + b(1+z)^e\left(1 + \exp{[(x-c)d]}\right)$ along with the best-fitting parameters. With this simple model, we can reproduce all of these predictions very efficiently, simply by picking the right set of parameters.  We find that the bias for the continuum bands is nearly wavelength-independent due to the mixing of star-forming and quiescent galaxies in a magnitude limited survey. 

We also compare our 5-parameter model with existing measurements from large scale structure surveys and forecasts, demonstrating that our model is in reasonable agreement with HIPASS, ALFAFA, EUCLID and 6dFGS within $1\sigma$ statistical uncertainties; the limit of our model on the faint end of the selection criteria arises when compared with B18. Future work could improve on this analysis by: 1) improving the understanding and modelling of the turn-down and upturn  features, 2) modelling the bias simultaneously as a function of magnitude/luminosity and color, 3) testing the dependence on cosmology and galaxy formation modelling. Notwithstanding the above improvements, this work provides an overview of the impact of galaxy physics on the bias, and allows for a quick estimation of the bias in a number of current or proposed large scale structure surveys.

\section*{Acknowledgements}

%The Acknowledgements section is not numbered. Here you can thank helpful colleagues, acknowledge funding agencies, telescopes and facilities used etc. Try to keep it short.

% Paragraph
We would like to thank the anonymous referee for a constructive report that has improved this paper. We acknowledge the discussion with Zheng Zheng, Xi Kang, Guoliang Li, Cedric Lacey and the simulations provided by Chris Power.
This work was supported by the UCAS Joint PhD Training Program, China Scholarship Council, and the Science and Technology Facilities Council of the United Kingdom.

This work used the DiRAC@Durham facility managed by the Institute for Computational Cosmology on behalf of the STFC DiRAC HPC Facility (www.dirac.ac.uk). The equipment was funded by BEIS capital funding via STFC capital grants ST/P002293/1, ST/R002371/1 and ST/S002502/1, Durham University and STFC operations grant ST/R000832/1. DiRAC is part of the National e-Infrastructure.

This work has benefited from the publicly available programming language {\sc python} (\url{https://www.python.org/}) and the package {\sc matplotlib} (\url{https://matplotlib.org/}).

%%%%%%%%%%%%%%%%%%%%%%%%%%%%%%%%%%%%%%%%%%%%%%%%%%



\bibliographystyle{mnras}
\bibliography{reference} % if your bibtex file is called example.bib

%\begin{thebibliography}{99}
%\end{thebibliography}

%%%%%%%%%%%%%%%%%%%%%%%%%%%%%%%%%%%%%%%%%%%%%%%%%%

%%%%%%%%%%%%%%%%% APPENDICES %%%%%%%%%%%%%%%%%%%%%

\begin{appendices}
\appendix
%\addtocontents{toc}{\protect\setcounter{tocdepth}{0}}
\section{UNCERTAINTIES OF LARGE-SCALE BIAS THROUGH ERROR PROPAGATION}
\label{appendixunc}
% Paragraph
Following the derivation of Gaussian covariance matrices of power spectrum from equation \eqref{eq:error_ps}, we derive the uncertainties of large-scale bias by error propagation as follows
\begin{equation}
	\begin{aligned}
        \sigma_{b_g(k)}^2 &\approx (\frac{\partial b_g}{\partial P_{g}})^2\sigma_{P_{g}}^2+(\frac{\partial b_g}
        {\partial P_{m}})^2\sigma_{P_{m}}^2+2\frac{\partial b_g}{\partial P_{g}}\frac{\partial b_g}
        {\partial P_{m}}\sigma_{P_{g}P_{m}}\\
        &=\frac{\sigma_{P_{g}}^2}{4P_{g}P_{m}} + \frac{P_{g}\sigma_{P_{m}}^2}{4P_{m}^3} - 
        \frac{\sigma_{P_{g}P_{m}}}{2P_{m}^2} \\
        &=\frac{b_g^2}{2n_{\rm modes}}(2-2r_g^2+\frac{2}{\bar{n}_gP_{g}}+\frac{2}
        {\bar{n}_mP_{m}}+\frac{1}{\bar{n}_g^2P_{g}^2}+\frac{1}{\bar{n}_m^2P_{m}^2})\\
        &=\frac{1}{n_{\rm modes}\bar{n}_g P_{m}},
        \label{eq:error_b}
	\end{aligned}
\end{equation}
where $P_{g}$ and $P_{m}$ refers to the power spectrum of galaxy and dark matter;  $\sigma_{P_g}=\frac{2}{n_{\rm modes}}\left(P_g+\frac{1}{\bar{n}_g}\right)^2$ and $\sigma_{P_m}=\frac{2}{n_{\rm modes}}\left(P_m+\frac{1}{\bar{n}_m}\right)^2$ according to equation~\eqref{eq:error_ps}; the $\sigma_{P_{g}P_{m}}= \frac{2}{n_{\rm modes}}P_{gm}^2$ \citep{2009MNRAS.397.1348W} and $r_g = P_{gm}/\sqrt[]{P_{g}P_{m}}\approx 1$ \citep{pen1998reconstructing,dekel1999stochastic,seljak2004large,bonoli2009halo}. Here we ignore the symbol (k) in the derivation for simplicity.

% Paragraph
We verified the theory and derivation using a set of 500 approximate halo catalogues. Dark matter simulations of size $L_{\rm box}$=512 $h^{-1}$Mpc with the number of particles $1024^{3}$ were generated using the approximate simulation code $\textsc{L-PICOLA}$ \citep{howlett2015picola}. Halos were then identified using a 3D FOF algorithm. For halo mass bins, we compute the power spectrum of the 500 simulations, and used these to estimate the variance. %The bias and variance in the same mass bins was estimated by taking the ratio between the halo and dark matter power spectra for each realisation. 

\begin{figure}
	\includegraphics[width=\columnwidth]{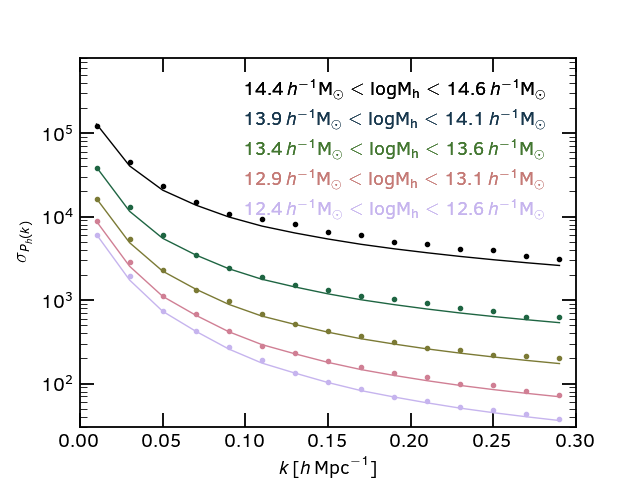}
    \includegraphics[width=\columnwidth]{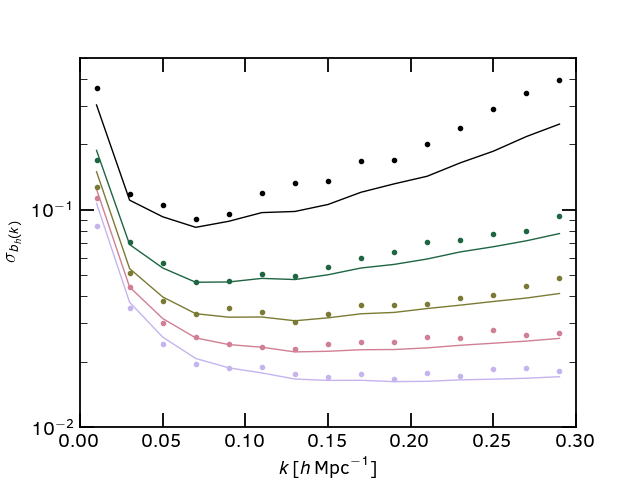}
    \caption{Upper: The error of power spectrum as a function of wavenumber $k$. 
    		 Lower: The error of halo bias as a function of wavenumber $k$. 
             The points and lines indicate measurements and theory respectively. 
             They are both color-coded by halo mass as shown on the upper panel.}
    \label{fig:sig}
\end{figure}

Fig.~\ref{fig:sig} shows the measured variance in the halo power spectrum and bias compared to theory. As seen, the points measured from the simulation are in good agreement with the lines i.e. equation \eqref{eq:error_ps} and \eqref{eq:berror}, over the most of mass ranges, but in the high mass bin, the equation \eqref{eq:error_b} intends to underestimate the noises where the wavenumber $\mathrm{k > 0.1 h/MPc}$ . This mismatch could be due to: 1) non-Poissonian shot-noise; 2) the fact we only used 500 simulations to measure this, thus there is some inaccuracy in the measurements which is difficult to quantify. Though these changes to our error formulae have little effect on the final fitting results.

% Figure

\section{OBSERVATIONAL CONSTRAINTS FOR \textsc{GALFORM}}
\label{appendixobs}

\begin{figure*}
	\includegraphics[width=2\columnwidth]{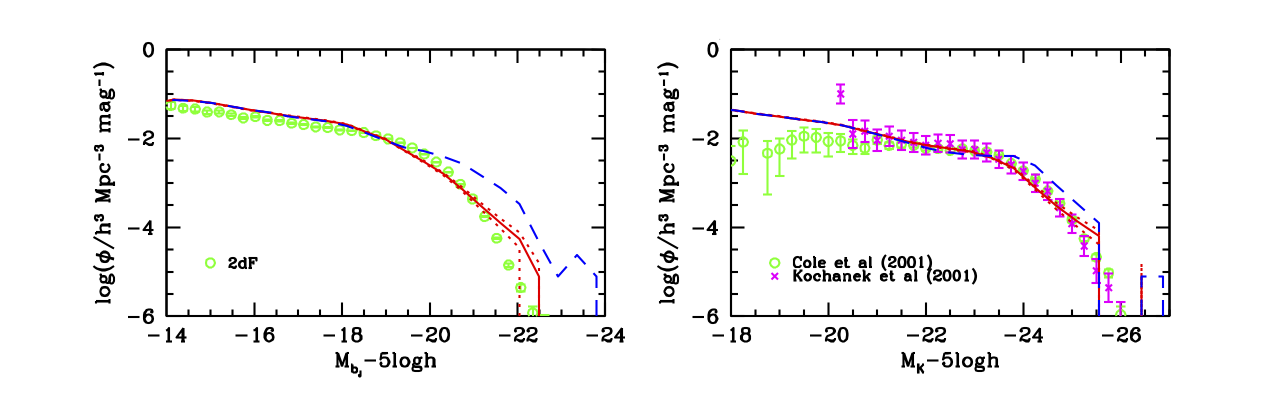}
    \caption{ The rest-frame $b_j$-band and K-band luminosity functions at z = 0. The solid and dashed lines show the results with and without duct extinction respectively. The dotted lines show the poisson errorbars for \textsc{GALFORM}. Observational data are from \citet{norberg20022df}, \citet{cole20012df} and \citet{kochanek2001k}.}
    \label{fig:bj}
\end{figure*}

\begin{figure*}
	\includegraphics[width=2\columnwidth]{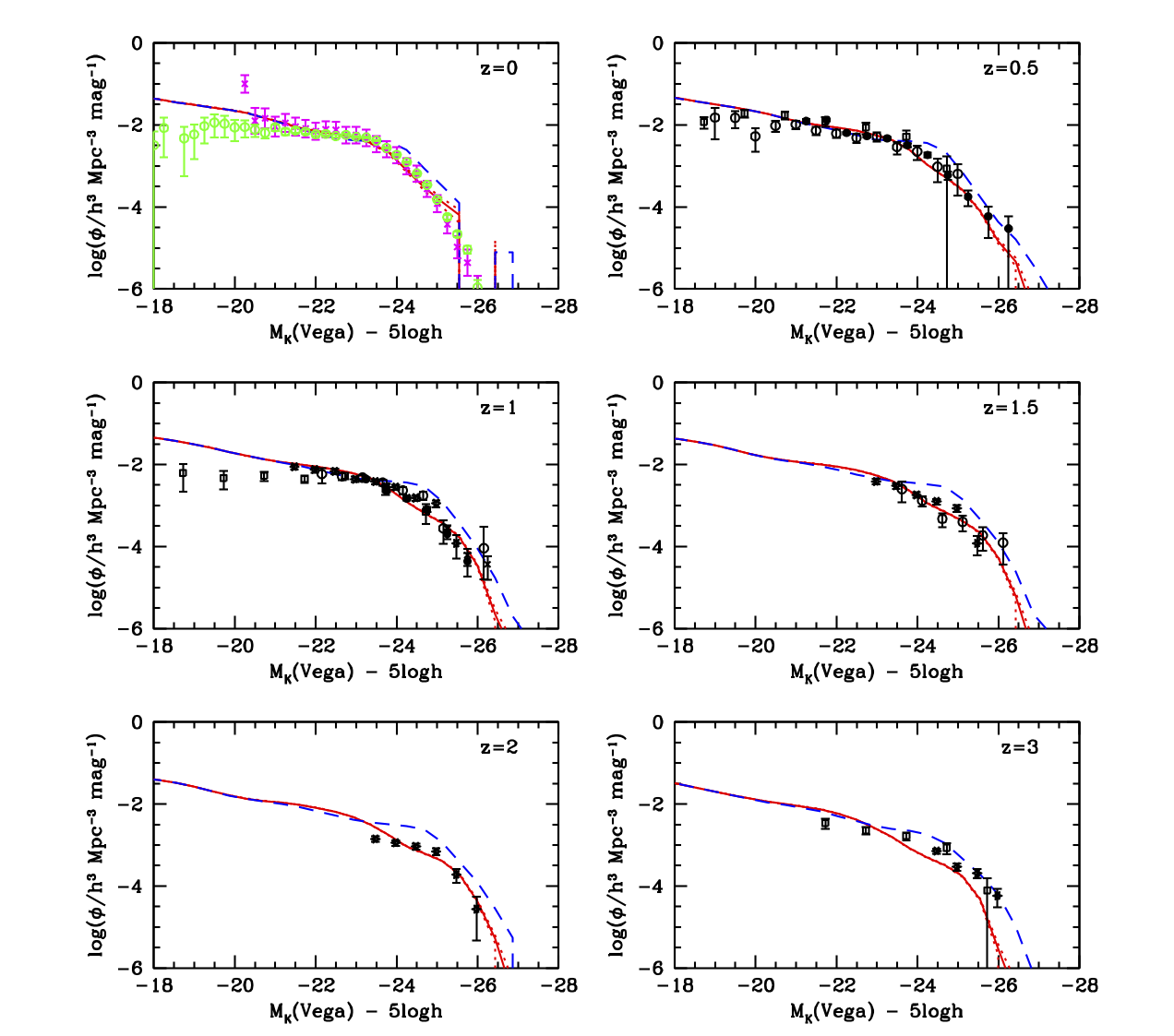}
    \caption{As in the right panel of Fig.~\ref{fig:bj}, but showing the evolution of the K-band luminosity function up to z = 3, as labelled. Observational data are from \citet{pozzetti2003k20} (open circles), \citet{drory2003munich} (crosses), \citet{saracco2006probing} (squares), \citet{caputi2006further} (hexagrams), \citet{cirasuolo2010new} (filled circles).
    }
    \label{fig:kband}
\end{figure*}

Fig.~\ref{fig:bj} shows the rest-frame $b_j$-band and K-band luminosity functions compared with observations at z = 0 in the local Universe. In Fig.~\ref{fig:kband}, we show the evolution of the rest-frame K-band luminosity function from z = 0 to z = 3. The solid lines show the predictions including dust extinction, while the dashed lines ignore the effects of dust extinction. As shown, our galaxy formation model is in reasonable agreement with the observations. Note that we are using slightly different captions for axes in this part from the main text to distinguish the rest-frame and observer-frame magnitudes.

% Figure

% Figure

\end{appendices}
%%%%%%%%%%%%%%%%%%%%%%%%%%%%%%%%%%%%%%%%%%%%%%%%%%

% Don't change these lines
\bsp	% typesetting comment
\label{lastpage}
\end{document}